\documentclass[showpacs,preprintnumbers,twocolumn,superscriptaddress,floatfix]{revtex4}
\usepackage{amsmath}
\usepackage{amsfonts}
\usepackage{amssymb}
\usepackage{graphicx}
\usepackage{array}
\usepackage{tabulary}
\usepackage{multirow}
\usepackage{subfigure}
\usepackage{epsfig}

\begin{document}

\newcommand{\mpt}{$\langle \mathrm{p_{T}} \rangle$}
\newcommand{\pt}{$\mathrm{p_{T}}$}
\newcommand{\kt}{$\mathrm{k_{T}}$}
\newcommand{\mt}{$\mathrm{m_{T}}$}
\newcommand{\Ks}{$\mathrm{K^{0}_{S}}$}
\newcommand{\kp}{$\mathrm{K^{+}}$}
\newcommand{\km}{$\mathrm{K^{-}}$}
\newcommand{\Kpm}{$\mathrm{K^{\pm}}$}
\newcommand{\pip}{$\mathrm{\pi^{+}}$}
\newcommand{\pim}{$\mathrm{\pi^{-}}$}
\newcommand{\dedx}{$dE/dx$}

\newcommand{\sps}{S$p\bar{p}$S}
\newcommand{\midy}{$\mid$y$\mid <$}
\newcommand{\dndy}{\ensuremath{dN/dy}}
\newcommand{\pbarp}{$p+\bar{p}$}
\newcommand{\sqs}{$\sqrt{\text{s}}$}

\title{Strange Particle Production in $p+p$ Collisions at $\sqrt{s}$= 200 GeV}

\affiliation{Argonne National Laboratory, Argonne, Illinois 60439}
\affiliation{University of Birmingham, Birmingham, United Kingdom}
\affiliation{Brookhaven National Laboratory, Upton, New York 11973}
\affiliation{California Institute of Technology, Pasadena, California 91125}
\affiliation{University of California, Berkeley, California 94720}
\affiliation{University of California, Davis, California 95616}
\affiliation{University of California, Los Angeles, California 90095}
\affiliation{Carnegie Mellon University, Pittsburgh, Pennsylvania 15213}
\affiliation{University of Illinois, Chicago}
\affiliation{Creighton University, Omaha, Nebraska 68178}
\affiliation{Nuclear Physics Institute AS CR, 250 68 \v{R}e\v{z}/Prague, Czech Republic}
\affiliation{Laboratory for High Energy (JINR), Dubna, Russia}
\affiliation{Particle Physics Laboratory (JINR), Dubna, Russia}
\affiliation{University of Frankfurt, Frankfurt, Germany}
\affiliation{Institute of Physics, Bhubaneswar 751005, India}
\affiliation{Indian Institute of Technology, Mumbai, India}
\affiliation{Indiana University, Bloomington, Indiana 47408}
\affiliation{Institut de Recherches Subatomiques, Strasbourg, France}
\affiliation{University of Jammu, Jammu 180001, India}
\affiliation{Kent State University, Kent, Ohio 44242}
\affiliation{Institute of Modern Physics, Lanzhou, China}
\affiliation{Lawrence Berkeley National Laboratory, Berkeley, California 94720}
\affiliation{Massachusetts Institute of Technology, Cambridge, MA 02139-4307}
\affiliation{Max-Planck-Institut f\"ur Physik, Munich, Germany}
\affiliation{Michigan State University, East Lansing, Michigan 48824}
\affiliation{Moscow Engineering Physics Institute, Moscow Russia}
\affiliation{City College of New York, New York City, New York 10031}
\affiliation{NIKHEF and Utrecht University, Amsterdam, The Netherlands}
\affiliation{Ohio State University, Columbus, Ohio 43210}
\affiliation{Panjab University, Chandigarh 160014, India}
\affiliation{Pennsylvania State University, University Park, Pennsylvania 16802}
\affiliation{Institute of High Energy Physics, Protvino, Russia}
\affiliation{Purdue University, West Lafayette, Indiana 47907}
\affiliation{Pusan National University, Pusan, Republic of Korea}
\affiliation{University of Rajasthan, Jaipur 302004, India}
\affiliation{Rice University, Houston, Texas 77251}
\affiliation{Universidade de Sao Paulo, Sao Paulo, Brazil}
\affiliation{University of Science \& Technology of China, Hefei 230026, China}
\affiliation{Shanghai Institute of Applied Physics, Shanghai 201800, China}
\affiliation{SUBATECH, Nantes, France}
\affiliation{Texas A\&M University, College Station, Texas 77843}
\affiliation{University of Texas, Austin, Texas 78712}
\affiliation{Tsinghua University, Beijing 100084, China}
\affiliation{Valparaiso University, Valparaiso, Indiana 46383}
\affiliation{Variable Energy Cyclotron Centre, Kolkata 700064, India}
\affiliation{Warsaw University of Technology, Warsaw, Poland}
\affiliation{University of Washington, Seattle, Washington 98195}
\affiliation{Wayne State University, Detroit, Michigan 48201}
\affiliation{Institute of Particle Physics, CCNU (HZNU), Wuhan 430079, China}
\affiliation{Yale University, New Haven, Connecticut 06520}
\affiliation{University of Zagreb, Zagreb, HR-10002, Croatia}

\author{B.I.~Abelev}\affiliation{Yale University, New Haven, Connecticut 06520}
\author{J.~Adams}\affiliation{University of Birmingham, Birmingham, United Kingdom}
\author{M.M.~Aggarwal}\affiliation{Panjab University, Chandigarh 160014, India}
\author{Z.~Ahammed}\affiliation{Variable Energy Cyclotron Centre, Kolkata 700064, India}
\author{J.~Amonett}\affiliation{Kent State University, Kent, Ohio 44242}
\author{B.D.~Anderson}\affiliation{Kent State University, Kent, Ohio 44242}
\author{M.~Anderson}\affiliation{University of California, Davis, California 95616}
\author{D.~Arkhipkin}\affiliation{Particle Physics Laboratory (JINR), Dubna, Russia}
\author{G.S.~Averichev}\affiliation{Laboratory for High Energy (JINR), Dubna, Russia}
\author{Y.~Bai}\affiliation{NIKHEF and Utrecht University, Amsterdam, The Netherlands}
% ...

\collaboration{STAR Collaboration}\noaffiliation

\date{\today}

\begin{abstract}
We present strange particle spectra and yields measured at mid-rapidity 
in $\sqrt{\text{s}}=200$ GeV proton-proton ($p+p$) collisions at RHIC. 
We find that the previously observed universal transverse mass ($\mathrm{m_{T}}\equiv\sqrt{\mathrm{p_{T}}^{2}+\mathrm{m}^{2}}$) scaling
of hadron production in $p+p$ collisions seems to break down at higher 
\mt~and that there is a difference in the shape of the \mt~spectrum 
between baryons and mesons.  We observe mid-rapidity anti-baryon to baryon ratios near 
unity for $\Lambda$ and $\Xi$ baryons and no dependence of the ratio 
on transverse momentum, indicating that our data do not yet reach the quark-jet  
dominated region.  We show the dependence of the mean transverse momentum (\mpt) 
on measured charged particle multiplicity and on particle mass and infer that these 
trends are consistent with gluon-jet dominated particle production.  
The data are compared to previous measurements from CERN-SPS, 
ISR and FNAL experiments and to Leading Order (LO) and Next to Leading 
order (NLO) string fragmentation model predictions.  We infer from these 
comparisons that the spectral shapes and particle yields from $p+p$ collisions 
at RHIC energies have large contributions from gluon jets rather than quark jets.
\end{abstract}


Version 4.2 \date{\today}

\maketitle

\section{Introduction \label{intro}}
The production of particles in elementary proton-proton ($p+p$) collisions
is thought to be governed by two mechanisms.  Namely, soft, thermal-like
processes which populate the low momentum part of the particle
spectra (the so-called underlying event) and the hard
parton-parton interaction process.  In this scenario, the low transverse momentum (\pt) 
part of the spectrum is exponential in transverse mass (\mt$\equiv\sqrt{\mathrm{m}^{2}+\mathrm{p_{T}}^{2}}$) 
while fragmentation, in leading order models, introduces a power law tail at high \pt.  
We investigate the validity of these assumptions at RHIC energies
by studying the spectral shapes and the yields of identified
strange hadron spectra from the lightest strange mesons (\Kpm) to
the heavy, triply-strange $\Omega^{-}$ baryon.

In this paper we report 
the results for transverse momentum spectra and mid-rapidity yields ($dN/dy$) of \Kpm, \Ks, $\Lambda$, 
$\overline{\Lambda}$, $\Xi^{-}$, $\overline{\Xi}^{+}$, and $\Omega^{-}+\overline{\Omega}^{+}$ measured 
by the STAR experiment during the 2001-2002 $\sqrt{\mathrm{s}}$=200 GeV $p+p$ running at RHIC. After 
a brief introduction in Section \ref{Exp} of the experimental setup and the conditions for this run, 
we provide a description of the event selection criteria and the efficiency of reconstructing the primary 
interaction vertex in Section \ref{Events}.  Specific attention will be given to the complications introduced by more 
than one event occurring in the detector during readout, a condition referred to as ``pile-up".  The details 
of strange particle reconstruction and the efficiency thereof will be discussed in Sections \ref{Reco}, \ref{Extraction}, 
and \ref{Eff}.  In Section \ref{Spectra} we describe the final measured \pt~spectra and 
mid-rapidity yields.  We also describe the functions that were used to parameterize the \pt~spectra in 
order to extrapolate the measurement to zero \pt.  We will show that the previously widely used power-law
extrapolation for $p+p$ and $p+\overline{p}$ collisions \cite{Bocquet96} does not yield the best 
$\chi^2$ results for the strange baryons and we will consider alternatives.
Section \ref{mt-scalingSection} introduces the idea of transverse mass scaling (\mt-scaling) 
and its applicability to our data.  The measured anti-particle to particle ratios are presented in Section 
\ref{Ratios}.
Interesting trends of increasing mean transverse momentum, \mpt, 
with particle mass have been previously observed in $p+p$ collisions at ISR energies 
(20 $\leq\sqrt{\text{s}}\leq$ 63 GeV) \cite{Alper75}.  Mean transverse momentum has 
also been found to increase with event multiplicity in $p+\overline{p}$ collisions at Sp$\overline{\text{p}}$S 
($\sqrt{\text{s}}=630$ GeV) \cite{Bocquet96} and FNAL energies (300 GeV $\leq\sqrt{\text{s}}\leq$ 
1.8 TeV) \cite{E735,E735-2}.  
We will show the dependence of our \mpt~measurements on both particle mass and event multiplicity in Section 
\ref{meanPtSec}.  We discuss the details 
of the experimental errors and then compare our results in Section \ref{models} with several models that attempt to describe 
particle production in $p+p$ collisions via pQCD, string fragmentation, and mini-jets \cite{Gyulassy92}.  
We conclude in Section \ref{Summary} with a discussion of the major results and some remarks about 
future directions for the ongoing analyses.

\section{Experimental Setup \label{Exp}}

\begin{figure}[h]
\epsfig{figure=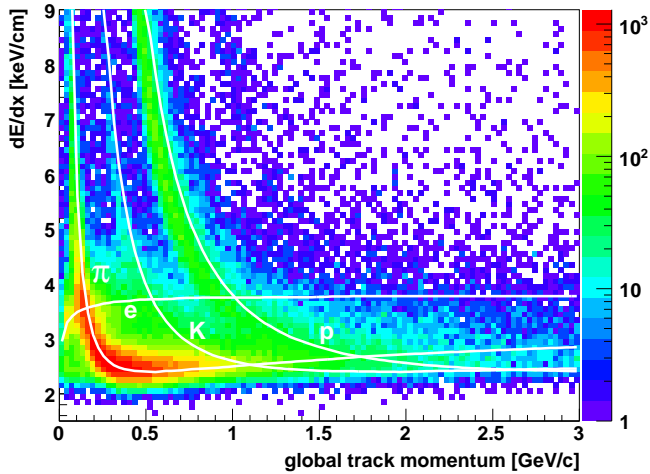,width=\columnwidth} 
\caption{\dedx~vs. Momentum for STAR $p+p$ collisions at $\sqrt{\mathrm{s}}$=200 GeV.  The curves 
are Bichsel parameterizations \cite{Bichsel}.} 
\label{fig:Bichsel}
\end{figure}

The data presented in this paper were collected with the STAR detector \cite{STAR}.  
The primary detector sub-system used for these analyses is the large cylindrical 
Time Projection Chamber (TPC), which is able to track charged
particles in the pseudo-rapidity range $|\eta| \leq 1.8$ with full azimuthal coverage \cite{TPC}.  The 
TPC has 45 pad rows in the radial direction allowing a maximum of 45 hits to be located on a 
given charged particle track.  A uniform magnetic field of 0.5 T is applied along the beam line by the 
surrounding solenoidal coils allowing the momentum of charged particles to be 
determined to within 2-7\% depending on the transverse momentum of the particle. 
The field polarity was reversed once during the 2001-2002 run to allow for 
studies of systematic errors.  The TPC tracking efficiency in $p+p$ collisions is greater than 90\% for 
charged particles with \pt~$\ge 300$ MeV/$c$ in
the pseudorapidity region $|\eta| < 0.7$ \cite{TPC}.  Particle 
identification may be achieved via measurements of energy loss due to 
specific ionization from charged particles passing through the TPC gas (\dedx). 
The \dedx, when plotted vs. rigidity separates the tracks into several bands which 
depend on the particle mass.  A semi-empirical formula describing the variation of 
\dedx~with rigidity is provided by the Bethe-Bloch equation \cite{Bichsel}.  An updated form, 
which accounts for the path length of a given particle through matter, has been given by Bichsel 
and provides a reasonable description of the \dedx~band centers for the particles presented in this 
paper \cite{Bichsel}.  The Bichsel curves are shown in Figure \ref{fig:Bichsel}.  

The dataset analyzed in this paper consisted of $1.4\times10^{7}$ minimally-biased events 
before cuts.  After applying a cut requiring the location of the primary vertex to 
be within 50 cm of the center of the TPC along the beam axis, to limit acceptance 
variations, $6\times10^{6}$ events remained.  In all events, the detectors were triggered by 
requiring the simultaneous detection of at least one charged particle at forward rapidities 
($3.5 \leq |\eta| \leq 5.0$) in Beam-Beam scintillating counters (BBCs) located at both 
ends of the TPC.  This is referred to as a minimally biased trigger. 
The BBCs are sensitive only to the non-singly diffractive (NSD) part (30 mb) of the $p+p$ 
total inelastic cross-section (42 mb) \cite{ppSpecPaper,STARBBCs}.  A more detailed 
description of STAR in general \cite{STAR} and the complete details 
of the TPC in particular \cite{TPC} can be found elsewhere.

\section{Analysis \label{Analysis}}

\subsection{Primary Vertex Finding and Event Selection \label{Events}}

The position of the interaction vertex is calculated by considering only those tracks which can 
be matched to struck slats of the STAR central trigger barrel (CTB) \cite{CTB}.
The CTB is a scintillating detector coarsely segmented into 240 slats placed azimuthally around 
the outside of the STAR TPC at a radius of 2 m.  It has a total pseudorapidity coverage 
of $-1.0<\eta<1.0$ and has a fast response time of 10-60 ns, which is roughly one 
quarter of the time between beam bunch crossings (218 ns in the 2001-2002 run).  
Therefore, in approximately 95.6\% of our 
$p+p$ collisions, only charged particles from the triggered event will produce signals in the CTB
which ensures that the primary vertex is initiated with tracks from the triggered event 
only (note that, unlike the BBCs, the CTB itself is not used as a trigger detector for the event 
sample presented here).  Furthermore, the primary vertex is assumed to be located somewhere 
along the known beam line.  The z co-ordinate (along the beam) of the primary vertex is then 
determined by minimizing the $\chi^2$ of the distance of closest approach of the tracks to 
the primary vertex.  

The RHIC beams were tuned so as to maximize the luminosity and, consequently, the 
number of collisions that can be recorded.  The average RHIC luminosities, which varied from 
$5\times10^{28}~\mathrm{cm^{-2}s^{-1}}$ to  $5\times10^{30}~\mathrm{cm^{-2}s^{-1}}$, 
produce collisions more frequently (on the order of 2-200 kHz) than the TPC can be read out (100 Hz).  
During $p+p$ running, as many as five pile-up events can overlap (coming in the $\sim$39 
$\mu$s before or after an event trigger) in the volume of the TPC.  
Pile-up events come earlier or later than the event trigger 
and tracks from pile-up events may therefore be only partially reconstructed as track fragments.  
These track fragments from a pile-up event 
can distort the determination of the location of the primary interaction vertex as 
they do not point back to the vertex of the triggered event.  
To solve this, tracks that do not match to a struck CTB slat are not used in the determination 
of the primary vertex position.
The remaining pile-up tracks, which match by chance to fired CTB slats, can then be removed with a 
reasonably restrictive (2-3 cm) analysis cut on a track's distance of closest approach 
to the determined primary vertex.

Another problem faced in the event reconstruction is the observation that for many minimally-biased 
triggers no primary vertex is reconstructed.  The problem is systematically worse
for the low multiplicity events.  Therefore, a correction must be applied to account
for the events that are triggered on yet lost in the analyses due to an
unreconstructed primary vertex.

The efficiency of the primary vertex finding software was
investigated by generating Monte Carlo (MC) $p+p$ events, propagating the
Monte Carlo produced particles through the STAR detector simulation
(GEANT), then adding the resulting simulated signals into the 
abort-gap events.  In an abort-gap event, the detectors are intentionally  
triggered when there are no protons in one or both of the beam bunches passing 
through the detector.  Abort-gap events therefore contain background due to 
the interaction of beam particles with remnant gas in the beam pipe and may 
also contain background remaining in the TPC from collisions in the crossings of previous or 
subsequent beam bunches.  Abort-gap events provide a realistic background 
environment in which to simulate the vertex finding process.  The embedded 
simulated event is then passed through the full software chain and
tracks are reconstructed.  These events are then compared to
the input from the MC events.  A quantity $\Delta(\text{z})$, representing 
the difference along the z (beam) axis between the actual embedded MC primary 
vertex and the reconstructed primary vertex is defined as follows:
\begin{equation}
\Delta(\text{z}) = \left| \mathrm{z}_{\text{PV}}^{\mathrm{MC}} - \mathrm{z}_{\text{PV}}^{\mathrm{reconstructed}} \right|.
\label{Eq:deltaPVz}
\end{equation}
\begin{figure}[h]
\epsfig{figure=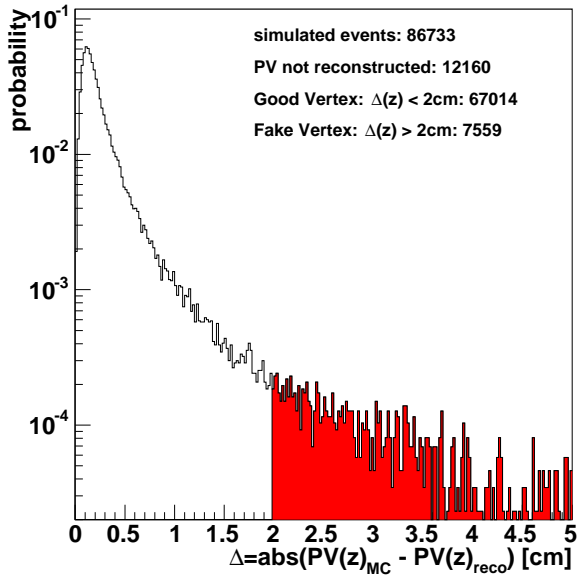,width=\columnwidth} 
\caption{Distribution of $\Delta(\text{z})$.  The unshaded region is the 
accepted range of good reconstructed event vetices.} 
\label{fig:deltaPVz}
\end{figure}

The probability distribution of $\Delta(\text{z})$ is shown in
Figure \ref{fig:deltaPVz} for approximately 87,000 simulated events.  
We separate events where the software finds a vertex into
two classes.  An event with a good primary vertex is defined as having $\Delta(\text{z})
\leq $ 2 cm, whereas a fake vertex event is one in which $\Delta(\text{z}) > $ 2 cm.
Whilst this limiting value is somewhat arbitrary, it does relate to offline cuts in
our particle reconstruction that are sensitive to the accuracy of the found vertex.
\begin{figure}[h]
\epsfig{figure=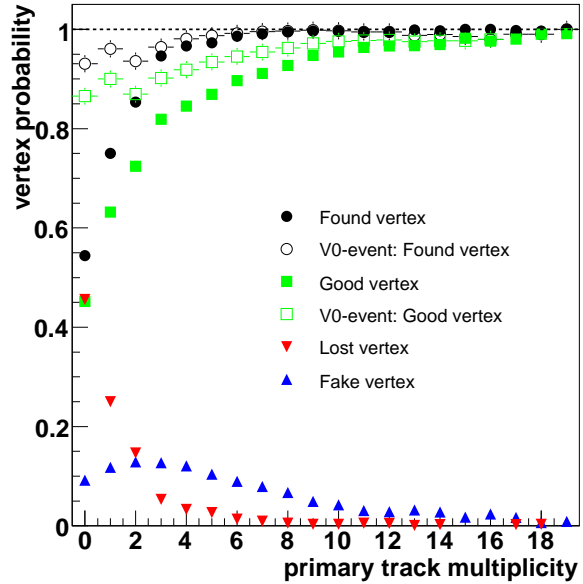,width=\columnwidth} 
\caption{Primary vertex finding efficiency vs measured primary track multiplicity. The 
horizontal line at one is only a guide for the eye.} 
\label{fig:PVeffy}
\end{figure}

It was found that the probability of finding the
primary vertex was strongly dependent on multiplicity.  For the purposes of this study,  
``charged track multiplicity'' is defined as being a count of tracks in the TPC  
that have at least 15 hits, at least 10 of which must be used in the track fit.  
After separating the raw charged track multiplicity distributions for each 
event class, \textit{i.e.} lost vertex, fake vertex and good vertex, 
these distributions can be divided by the charged track multiplicity distribution 
of all events.  This ratio then represents the
probability for a certain event class to occur as a function of the
measured charged track event multiplicity.  Finally, the probabilities for each charged 
track multiplicity are mapped back to the corresponding primary track multiplicity, 
where ``primary tracks'' are those which satisfy the above requirements and 
additionally point back to within 3 cm of the primary vertex.  The probabilities for 
each event class as a function of primary track multiplicity are shown in Figure
\ref{fig:PVeffy}.  Whereas lost vertex events are monotonically
decreasing with increasing multiplicity, fake vertex events are most
probable when the event has 2 primary tracks.  The open symbols in Figure 
\ref{fig:PVeffy} show the corresponding ``found'' and ``good'' 
probabilities for events that contained at least one strange particle decay candidate.  
Note that primary vertex finding is initiated with tracks pointing at fired slats of the 
CTB, as mentioned above.  But all found tracks are allowed to contribute to the final vertex 
position.  Therefore, on rare occasions and in low multiplicity events, a vertex may be 
found with no single track pointing back within 3 cm.  These events will appear in 
Figure \ref{fig:PVeffy} as having a found (or fake) vertex but zero primary track 
multiplicity.  The use of these probabilities to correct the strange particle yields 
and event counts as a function of multiplicity is described later.

\subsection{Particle Identification \label{Reco}}

All the strange particles presented here, with the exception of the charged kaons,  
were identified from the topology of their weak decay products in the dominant channel:
\begin{equation}
\mathrm{K^{0}_{S}} \rightarrow \pi^{+} + \pi^{-} \: (68.6\%) \label{Eq:K0Decay}
\end{equation}
\begin{equation}
\Lambda \rightarrow p + \pi^{-} \: (63.9\%) \label{Eq:LamDecay}
\end{equation}
\begin{equation}
\Xi^{-}  \rightarrow \Lambda + \pi^{-} \: (99.9\%) \label{Eq:XiDecay}
\end{equation}
\begin{equation}
\Omega^{-}  \rightarrow \Lambda + K^{-} \: (67.8\%) \label{Eq:OmegaDecay}
\end{equation}

The charged tracks of the daughters of neutral strange particle 
decays form a characteristic ``V"-shaped topological pattern known as a ``V0".  
The V0 finding software pairs oppositely charged particle tracks to form 
V0 candidates.  These candidates can then be further paired with a single 
charged track, referred to as the ``bachelor" to form candidates for 
$\Xi^{-}$ and $\Omega$ decays.  During the initial finding process, loose cuts 
are applied to partially reduce the background while maximizing the candidate pool.  
Once the candidate pool is assembled, a more stringent 
set of cuts is applied to maximize the signal-to-noise ratio and 
ensure the quality of the sample.  The cuts are analysis dependent and are 
summarized in Table \ref{tab:cutsV0} for the \Ks~and $\Lambda$ analyses and 
in Table \ref{tab:cutsXi} for the $\Xi$ and $\Omega$ analyses.
\begin{table}[ht]
\begin{center}
\setlength\extrarowheight{2pt}
\begin{tabular}{|c|c|c|c|c|c|c|}
\hline \textbf{Cut} & $\mathbf{\bf K^{0}_{S}}$ & $\mathbf{\Lambda}$ and $\mathbf{\overline{\Lambda}}$ \tabularnewline
\hline DCA of V0 to primary vertex  & $<$ 2.0 cm & $<$ 2.0 cm \tabularnewline
\hline DCA of V0-daughters   & $<$ 0.9 cm & $<$ 0.9 cm \tabularnewline
\hline N(hits) daughters  & $>$ 14 & $>$ 14 \tabularnewline
\hline N($\sigma$) \dedx  & $<$ 3 & $<$ 5 \tabularnewline
\hline Radial Decay Length       & $>$ 2.0 cm & $>$ 2.0 cm \tabularnewline
\hline Parent Rapidity (y)       & $\pm$ 0.5 & $\pm$ 0.5 \tabularnewline
\hline
\end{tabular} 
\caption{Summary of \Ks~and $\Lambda$ cuts.  See text for further explanation.}
\label{tab:cutsV0}
\end{center}
\end{table}
\begin{table}[h]
\begin{center}
\hspace*{-0.75cm}
\setlength\extrarowheight{2pt}
\begin{tabular}{|c|c|c|c|c|c|}
\hline \textbf{Cut} & $\mathbf{\Xi^{-}}$ and $\mathbf{\overline{\Xi}^{+}}$ & $\mathbf{\Omega^{-}}$ and $\mathbf{\overline{\Omega}^{+}}$\tabularnewline
\hline Hyperon Inv. Mass & 1321 $\pm$ 5 MeV & 1672 $\pm$ 5 MeV \tabularnewline
\hline Daughter $\Lambda$ Inv. Mass & 1115 $\pm$ 5 MeV & 1115 $\pm$ 5 MeV \tabularnewline
\hline N($\sigma$) \dedx~bachelor & $<$ 5 & $<$ 3 \tabularnewline
\hline N($\sigma$) \dedx~pos. daugh. & $<$ 5 & $<$ 3.5 \tabularnewline
\hline N($\sigma$) \dedx~neg. daugh. & $<$ 5 & $<$ 3.5 \tabularnewline
\hline N(hits) bachelor       & $>$ 14 & $>$ 14 \tabularnewline
\hline N(hits) pos. daugh.  & $>$ 14 & $>$ 14 \tabularnewline
\hline N(hits) neg. daugh.  & $>$ 14 & $>$ 14 \tabularnewline
\hline Parent Decay Length (lower) & $>$ 2.0 cm & $>$ 1.25 cm \tabularnewline
\hline Parent Decay Length (upper) & $<$ 20 cm & $<$ 30 cm \tabularnewline
\hline Daugh. V0 Decay Length (lower) & N/A & $>$ 0.5 cm \tabularnewline
\hline Daugh. V0 Decay Length (upper) & N/A & $<$ 30 cm \tabularnewline
\hline DCA of Parent to PVtx  & N/A & $<$ 1.2 cm \tabularnewline
\hline DCA of Daughters       & N/A & $<$ 0.8 cm \tabularnewline
\hline DCA of V0 Daughters    & N/A & $<$ 0.8 cm \tabularnewline
\hline DCA of Bachelor to PVtx (lower)  & N/A & $>$ 0.5 cm \tabularnewline
\hline DCA of Bachelor to PVtx (upper)  & N/A & $<$ 30 cm \tabularnewline
\hline Parent Rapidity        & $\pm$ 0.5 & $\pm$ 0.5 \tabularnewline
\hline
\end{tabular} 
\caption{Summary of $\Xi$ and $\Omega$ cuts.  See text for further explanation.}
\label{tab:cutsXi}
\end{center}
\end{table}

Several of these cuts require some further explanation.  A 
correlation has been observed between the luminosity and the 
raw V0 multiplicity.  This correlation is suggestive of 
pile-up events producing secondary V0s.  The apparent path 
of the V0 parent particle (the \Ks~or $\Lambda$) is extrapolated 
back towards the primary vertex.  The distance of closest approach
(DCA) of the V0 parent to the primary vertex is then determined. 
Secondary V0s from pile-up events do not point back well to the 
primary vertex of the triggered event and may therefore be removed 
via a cut on the DCA of the V0 parent to the primary vertex.  Parent particles 
for secondary V0s may be charged and curve away from the primary vertex 
before decaying, causing the secondary V0 to also point back poorly.  
Therefore, this cut also removes some true secondary V0s.

Tracks in the TPC are occasionally broken into two 
or more segments that appear to be independent tracks to the V0 
and $\Xi$ finding software.  In the majority of cases, this is due 
to tracks crossing the boundaries between sectors of the TPC pad plane.  
A cut requiring a minimum number of hits is applied to each of the 
decay daughter tracks to minimize the contamination from these track fragments.

\begin{figure*}[ht]
\epsfig{figure=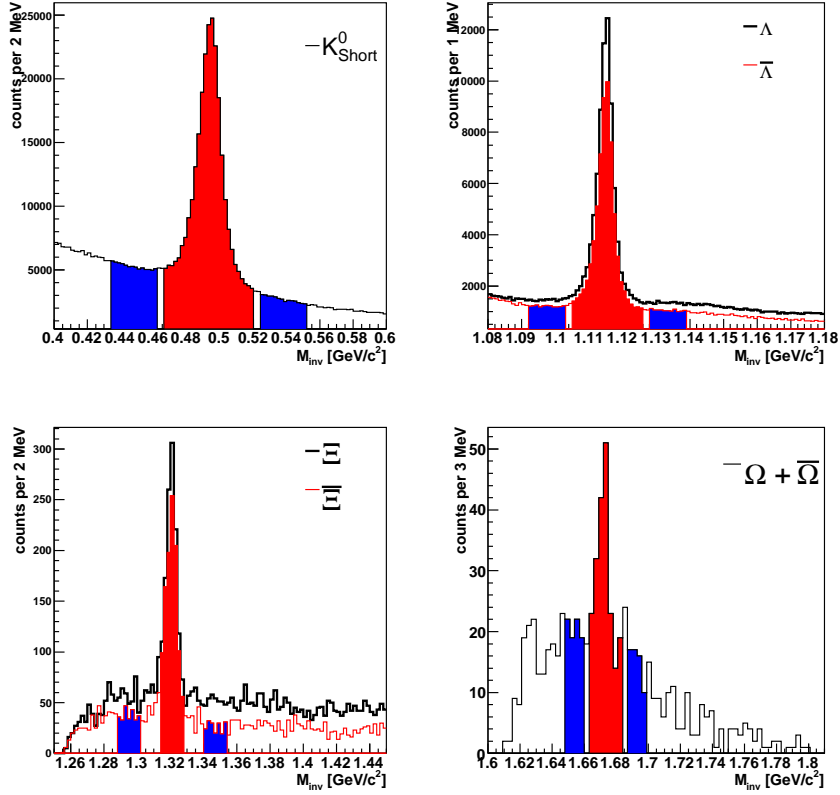,width=0.65\linewidth} 
\caption{Invariant mass distribution of
$\mathrm{K^{0}_{S}}$, $\Lambda$, $\overline{\Lambda}$, $\Xi^{-}$, 
$\overline{\Xi}^{+}$, and $\Omega^{-}+\overline{\Omega}^{+}$ after applying the geometrical cuts 
outlined in Tables \ref{tab:cutsV0} and \ref{tab:cutsXi}.} 
\label{fig:invmass}
\end{figure*}

Lastly, we define a variable N($\sigma$) to quantitatively 
measure the distance of a particular track to a certain particle band
in \dedx~vs. rigidity space \cite{STAR130Mult,STAR130p-bar-p}:
\begin{equation}
\mathrm{N}(\sigma) = \dfrac{dE/dx_{\mathrm{measured}} - dE/dx_{\mathrm{Bichsel}}}
{(\mathrm{R}/\sqrt{\mathrm{N_{samples}}}) \cdot dE/dx_{\mathrm{measured}}}
\end{equation}
where R is the \dedx~resolution (width in \dedx~of the distribution of a given 
particle band, see Fig. \ref{fig:Bichsel}) at the track's momentum and 
$\mathrm{N_{samples}}$ is the number of hits used in the determination of the \dedx.  
Cutting on the N($\sigma$) of a given track helps to decrease the background even 
further by decreasing the contamination of the candidate pool due to misidentified 
tracks.  This is particularly important for the $\Omega$ analysis.  The $\Lambda$ 
and $\Xi$ analyses can tolerate more open cuts in favor of increased statistics.  
The invariant mass distributions for \Ks, $\Lambda$, $\Xi^{-}$, $\Omega$ and their 
corresponding antiparticles are shown in Figure \ref{fig:invmass}.

The charged kaon decay reconstruction method is based on the fact that the 
four dominant \Kpm~decay channels (shown in relation \ref{Eq:KinkDecay}) have the 
same pattern.  The charged kaon decays into one or two neutral daughter(s) which 
are not detected and one charged daughter which is observed in the TPC.  
\begin{equation}
\mathrm{K^{\pm}} \rightarrow
\begin{cases}
                 \mu^{\pm} + \nu_{\mu} & (63.4\%) \\ 
                 \pi^{\pm} + \pi^{0} & (21.1\%) \\
                 \mu^{\pm} + \pi^{0} + \nu_{\mu} & (3.27\%) \\
                 \pi^{\pm} + \pi^{0} + \pi^{0} & (1.73\%)
\end{cases}
\label{Eq:KinkDecay}
\end{equation}

The decay topology corresponding to the above channels is known as a ``kink", as 
the track of the charged parent in the TPC appears to have a discontinuity at the 
point of the parent decay.  The kink-finding software starts by looping over all 
tracks reconstructed in the TPC in the given event, looking for pairs of tracks which 
are compatible with the kink pattern described above.  The first selection 
criterion is for the kaon decay vertex (the kink) to be found in a fiducial volume 
in the TPC.  The TPC has an inner radius of 50 cm and an outer radius of 200 cm from 
the nominal beamline, but the fiducial volume is defined to have an inner radius of 
133 cm and an outer radius of 179 cm.  The fiducial volume is chosen to 
suppress background due to high track densities (inner cut) while allowing a 
reasonable track length for the determination of the daughter momentum 
(outer cut).  This leads to a maximum number of hits for both the parent and 
daughter track in the fiducial volume.  Additional cuts are applied to the 
found track pairs in order to select the kink candidates.  
\begin{table}[ht]
\begin{center}
\setlength\extrarowheight{2pt}
\begin{tabular}{|>{\centering\hspace{0pt}}m{0.25\columnwidth}|>{\centering\hspace{0pt}}m{0.75\columnwidth}|}
\hline \textbf{Cut} & $\mathbf{\bf K^{\pm}}$ (kinks) \tabularnewline
\hline Invariant mass & $0.3 < \text{m}_{\text{inv}} < 1.0$ GeV/$c^2$ \tabularnewline
\hline Kink angle & $\begin{cases}
                      > \text{asin}(\mathrm{p}/\mathrm{M1})+4.-1.25 \cdot \mathrm{M1} & \text{for p$>$M1} \\
                      < \text{asin}(\mathrm{p}/\mathrm{M2}) & \text{for p$>$M2}
                      \end{cases}$ \tabularnewline
\hline Daughter mom. & $>$100 MeV/$c$ \tabularnewline
\hline {DCA/cm between Parent and Daughter} & $< 0.123 + 0.082/{({\mathrm{p_{T}/GeV}/c)}^{1.153}}$  \tabularnewline
\hline
\end{tabular} \\
\caption{Summary of cuts used in the kink analysis.  The notation is as follows: \\
$\mathrm{M1} \equiv (m_{\pi}^{2}-m_{\mu}^{2})/2m_{\mu}$ and $\mathrm{M2} \equiv (m_{\mathrm{K}}^{2}-m_{\mu}^{2})/2m_{\mu}$.  
See text and Figure \ref{fig:kaonCut} for further details. }
\label{tab:cutsKink}
\end{center}
\end{table} 
\begin{figure}[ht]
\epsfig{figure=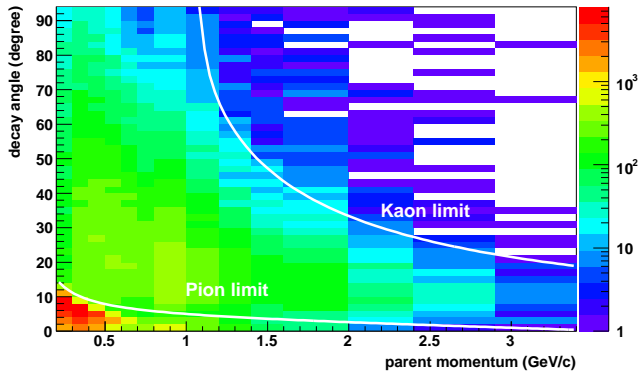,width=\columnwidth} 
\caption{Kink angle cut regions for \kp~and \km~identified via the kink method.} 
\label{fig:kaonCut}
\end{figure}

For each kink found, a mass hypothesis is given to both the parent and daughter 
tracks (i.e. \kp~parent and $\mu^{+}$ daughter) and the pair invariant mass is 
calculated based on this hypothesis.  A cut on the invariant mass 
($\text{m}_{\text{inv}}$ in Table III) can then be applied.  As charged pions 
decay with a branching ratio of approximately 100\% into the same $\mu+\nu_{\mu}$ 
channel as the charged kaons, they will have the same track decay topology in the TPC.  
We therefore expect that the kink finding algorithm described above will 
include \kp, \km, \pip~and \pim~as kink parent candidates.
Therefore, several other cuts must be applied to further eliminate 
the pion background from the kaon decays in which we are interested.   A summary of 
the applied cuts is given in Table \ref{tab:cutsKink}.

In Figure \ref{fig:kaonCut} we show the regions excluded by the kink angle cut in Table 
\ref{tab:cutsKink}.  The cut is placed on the line marking the pion limit.  In addition to 
the cuts listed in Table \ref{tab:cutsKink}, a cut was applied to the specific energy loss 
to remove pion contamination below \pt=500 MeV/$c$ where the kaon and pion \dedx~bands are 
clearly separated.

The parent-daughter DCA cut in the last row of Table \ref{tab:cutsKink} 
was determined from a two-dimensional (DCA and \pt) study of the background.  
The appropriate cut level was determined in each (DCA,\pt) cell and the 
results were fit with a function of the form $A+(B \cdot \mathrm{p_{T}}^{-C})$.  The 
resulting parameters $A$, $B$, and $C$ are given in Table \ref{tab:cutsKink}.

\subsection{Signal Extraction \label{Extraction}}

In order to extract the particle yield and \mpt, we build invariant 
mass distributions in several \pt~bins for each of the particle species except 
the charged kaons.  The residual background in each \pt~bin is then subtracted through 
a method referred to here as ``bin-counting".  

In the bin-counting method, three regions are defined in the invariant-mass 
distribution.  The first, which is defined using the Gaussian signal width found by fitting 
the \pt-integrated invariant-mass distribution with a linear function plus a Gaussian, is 
the region directly under the mass peak ($\pm 3.5\sigma$, $\pm 4.5\sigma$, and $\pm 2.5\sigma$ for the 
\Ks, $\Lambda$, and $\Xi$ respectively) which includes both signal and background 
(red or lightly shaded in Figure \ref{fig:invmass}).  For the \Ks~and $\Lambda$ invariant mass distributions, the second and third regions (blue or dark shading in Figure \ref{fig:invmass}) are defined to be the same total width as the signal region placed on either side ($1\sigma$ away for \Ks~and $\Lambda$) of the chosen signal region.  For the $\Xi$, the second and third regions are each the size of the 
signal region and are placed $4\sigma$ away.  In this method the background is implicitly taken to be linear under the mass peak.  In \pt~bins where the background appears to deviate significantly from the linear approximation, a second degree polynomial fit is used to determine the 
background under the mass peak.  This occurs mainly at low \pt.

This procedure is carried-out in each transverse momentum bin and as a function 
of event multiplicity.  The resulting spectrum is then corrected for vertex finding efficiency 
(Section \ref{Eff}) as well as the particle specific efficiency and acceptance (Section \ref{Events}).  
The $\Lambda$ and $\overline{\Lambda}$ spectra are further corrected for higher-mass feed-down as 
detailed in section \ref{Feeddown}.

\subsection{Particle Reconstruction Efficiencies \label{Eff}}

The number of reconstructed strange particles is less
than the actual number produced in the collision due to the 
finite geometrical acceptance of the detector and the efficiency 
of the tracking and decay-finding software.  Additionally,
the quality cuts described in section \ref{Reco} reduce not only the
combinatorial background but the raw signal as well.

\begin{figure*}[ht]
\epsfig{figure=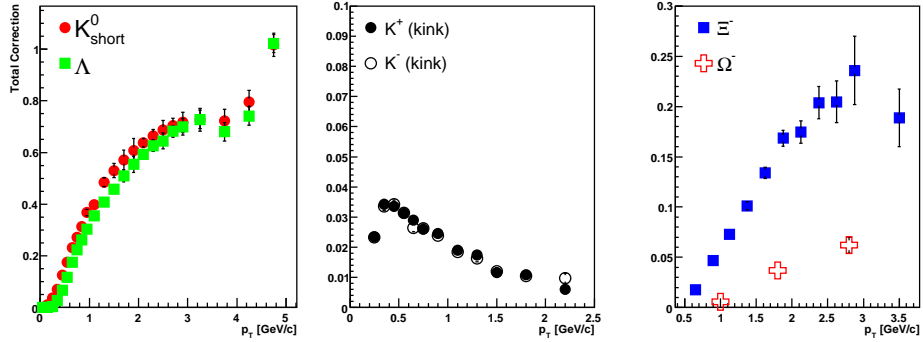,width=1.5\columnwidth} 
\caption{Total correction factor (efficiency $\times$ acceptance) for \Ks, $\Lambda$, 
\kp, \km, $\Xi$, and $\Omega$ after cuts} 
\label{fig:embedding}
\end{figure*}

In order to determine the efficiency for each particle species 
as a function of transverse momentum, an embedding process, similar to 
that described in section \ref{Events}, is employed.  In this process, 
a Monte Carlo generator is used to produce the particles of interest 
with a given transverse momentum distribution.  The produced particles 
are propagated through the GEANT detector simulation and the 
resulting signals embedded into real events at the level 
of the detector response (pixel level).  Using real events provides a 
realistic tracking and finding environment for evaluating the 
performance of the software.  Only one simulated particle is embedded 
in any given event so as not to overly modify the tracking and finding 
environment.  The embedded events are then 
processed with the full reconstruction software chain and the results 
compared with the input to determine the final correction factors for 
the transverse momentum spectra.  Whether or not the event used for 
embedding already contained one or more strange particles is not a 
concern as only GEANT-tagged tracks are counted for the purpose of 
calculating efficiencies.  The resulting total efficiencies 
(acceptance $\times$ tracking, finding, and cut efficiencies) are plotted in 
Figure \ref{fig:embedding} for \Ks, \Kpm, $\Lambda$, $\Xi$, and $\Omega$.
The correction is assumed to be constant over the measured rapidity region.

Finally, a correction needs to be applied to the raw particle yields
due to low primary vertex efficiencies for low multiplicity events
described in section \ref{Events}.  The spectra were binned in
multiplicity classes and for each class the particle yields were
corrected using the probabilities corresponding to finding a good vertex 
in an event with at least a V0 candidate (open squares in Figure \ref{fig:PVeffy}), 
thereby accounting for particles from lost and fake events.  The overall 
event normalization is also corrected, using the numbers corresponding to 
the probability of finding a vertex (black filled circles in Figure 
\ref{fig:PVeffy}), to account for the number of lost events.

\begin{table*}[ht]
\begin{center}
\setlength\extrarowheight{2pt}
\begin{tabular}{|c|c|c|c|c|c|c|c|c|c|c|}
\hline & \multicolumn{2}{c}{$\mathbf{\bf K^{0}_{S}}$}\vline &
         \multicolumn{2}{c}{$\mathbf{\bf K^{\pm}}$ (kinks)}\vline & 
         \multicolumn{2}{c}{$\mathbf{\Lambda (\overline{\Lambda})}$}\vline & 
         \multicolumn{2}{c}{$\mathbf{\Xi (\overline{\Xi})}$}\vline & 
         \multicolumn{2}{c}{$\mathbf{\Omega+\overline{\Omega}}$}\vline \tabularnewline 
\hline \textbf{Error Source} & 
       \textbf{dN/dy} & $\mathbf{\bf \langle p_{T} \rangle}$ & 
       \textbf{dN/dy} & $\mathbf{\bf \langle p_{T} \rangle}$ & 
       \textbf{dN/dy} & $\mathbf{\bf \langle p_{T} \rangle}$ & 
       \textbf{dN/dy} & $\mathbf{\bf \langle p_{T} \rangle}$ &
       \textbf{dN/dy} & $\mathbf{\bf \langle p_{T} \rangle}$ \tabularnewline
\hline Cuts and Corrections (\%)           & 5.4 & 1.1 &       
                                         3.7 & 2.2 &        
                                         5.4 & 1.3 &       
                                         13 & 1.1 &       
                                         15 & 8.0 \tabularnewline                
\hline Yield extraction and Fit function (\%) & 4.9 & 3.7 &   
                                                1.5 & 1.2 &        
                                                6.3 & 4.7 &   
                                                30 & 5.6 &  
                                                20\footnote{The numbers for $\Omega+\overline{\Omega}$ are 
for yield extraction only.  The statistics do not allow a meaningful fit function 
study to be done for $\Omega+\overline{\Omega}$.}\label{ftnote:omega} & 3.0\footnotemark[\value{mpfootnote}] \tabularnewline           
\hline Normalization (\%)   & 4 & N/A &   
                              4 & N/A &   
                              4 & N/A &   
                              4 & N/A &   
                              4 & N/A \tabularnewline      
\hline
\end{tabular}
\caption{A summary of systematic errors from various sources.  Errors from yield 
extraction and fit function for $\Xi$ are from comparison between \mt-exponential 
and power-law fits.  The normalization error affects only the particle yields.  
All entries are percentages.} 
\label{tab:SystErr}
\end{center}
\end{table*}

\subsection{Feed-down corrections \label{Feeddown}}

$\Xi$ and $\Omega$ baryons produce a $\Lambda$ as 
one of their decay products.  In some cases, the daughter $\Lambda$ 
can be detected as if it were a primary 
$\Lambda$ particle.  The result is a modification of the measured primary 
$\Lambda$ \pt~spectrum and an overestimation of the primary $\Lambda$ yield.  
The amount of contamination is unique to the cuts used to find the $\Lambda$.  
\begin{figure}[ht]
\epsfig{figure=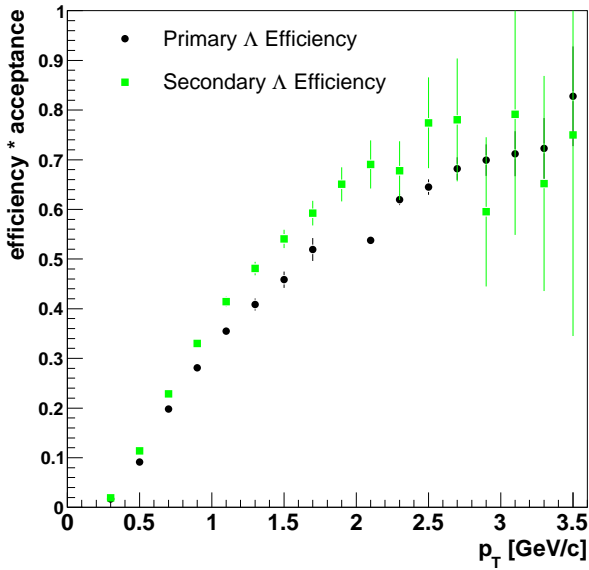,width=\columnwidth} 
\caption{Efficiency times acceptance for primary $\Lambda$s and $\Lambda$s 
coming from $\Xi$ decays.} 
\label{fig:feeddown}
\end{figure}

In order to correct this, Monte-Carlo $\Xi$ simulations were performed and tuned 
to match the measured shape and yield of the $\Xi$ \pt~spectrum presented in this paper.  
The shape and yield of the $\Lambda$ spectrum coming from $\Xi$ decays can then 
be determined.  The total correction factor (efficiency $\times$ acceptance) 
was then calculated for both primary $\Lambda$ baryons and secondary $\Lambda$ baryons 
produced by embedded $\Xi^{-}$ decays (see Figure \ref{fig:feeddown}).  The correction 
factor is different for $\Lambda$ baryons coming from $\Xi$ decays.  Lastly, the 
secondary $\Lambda$ spectrum is multiplied by the correction factor for secondary $\Lambda$ baryons, 
divided by the primary $\Lambda$ correction factor, and the result is subtracted from the measured 
$\Lambda$ spectrum.  The application of the correction factor is formalized in Equation \ref{equ:FDcorrection}, 
\begin{equation}
FD(\mathrm{p_{T}}) = Measured(\mathrm{p_{T}}) - (MC(\mathrm{p_{T}}) \cdot R_{eff}(\mathrm{p_{T}}))
\label{equ:FDcorrection}
\end{equation}
where $FD(\mathrm{p_{T}})$ is the final feed-down corrected \pt~spectrum, $Measured(\mathrm{p_{T}})$ is 
the non-feed-down corrected \pt~spectrum (corrected for efficiency and acceptance), $MC(\mathrm{p_{T}})$ is 
the secondary \pt~spectrum (determined from MC), and $R_{eff}(\mathrm{p_{T}})$ is the ratio of the 
secondary efficiency and acceptance correction to the primary efficiency and acceptance correction.  
The neutral $\Xi^{0}$ has not been measured by our 
experiment and therefore, for the purposes of determining the feed-down correction, 
the $\Xi^{0}$ yield is taken to be equal to the measured $\Xi^{-}$ yield.  Similarly, 
the $\overline{\Xi}^{0}$ yield is taken to be equal to the measured 
$\overline{\Xi}^{+}$ yield.  By using Monte Carlo calculations, we determined that 
the finding efficiency for secondary $\Lambda$ particles was the 
same whether the $\Lambda$ comes from a charged or neutral $\Xi$.  Therefore, 
the final feed-down correction is doubled to account for feed-down from $\Xi^{0}$ 
decays.

\subsection{Systematic Errors}

Several sources of systematic errors were identified in the
analyses.  A summary of these errors and their estimated size is to
be found in Table \ref{tab:SystErr}.  A description of various sources of
systematic error and their relative contribution is given below. 

\textbf{Cuts and Corrections:} The offline cuts that are applied to minimize the residual 
backgrounds also help eliminate contamination from pile-up events.  The 
cuts may be tightened to further reduce background or loosened to allow 
more signal and improved statistics at the cost of greater contamination.  
The final cuts are a compromise between these two extremes which aim to 
maximize the statistics for a given particle species while eliminating 
as much background as possible.  The systematic errors from the
cut-tuning provide an estimate for our sensitivity to changes in the 
various cuts.

This number includes the systematic errors from the 
embedding and vertex finding efficiency corrections.  The $\Lambda$ and 
$\overline{\Lambda}$ entry also accounts for the systematic errors from the feed-down correction.

\textbf{Methods of Yield Extraction:} In order to estimate the systematic 
error on the yield extraction in each \pt~bin, a second method of determining 
the yield in a given bin was used.  In the second method, a combination of Gaussian
plus a linear function is fit to the mass peak and background.  The yield is then determined 
by subtracting the integral of the fitted linear function across the width of the signal peak from 
the sum of the bin content in the peak.  In both methods, fitting and bin-counting, the background 
is assumed to be linear under the mass peak.  A second degree polynomial fit is used 
in \pt~bins where this assumption is clearly invalid (mostly at low \pt).  The two methods 
of extracting the yield may give different values due to the finite precision of the 
fitting method and fluctuations in the background in the bin-counting method.  The 
difference in the two methods and any differences resulting from a deviation from 
the linear background assumption are taken into account by this systematic error.

The systematic error on the lowest $\chi^{2}$/ndf functional parameterization of the 
spectra is also included in this row.  It is estimated by comparing the yield and 
\mpt~from the best $\chi^{2}$/ndf functional parameterization with that from the second 
best $\chi^{2}$/ndf functional form.  The final numbers for the mid-rapidity yield and 
\mpt~(in Tables \ref{tab:yields} and \ref{tab:meanpt}) were determined for each particle 
using the fit with the smallest $\chi^{2}$/ndf as shown in Table \ref{tab:chi2}.

\textbf{Normalization:} The overall systematic error from the vertex and
trigger efficiency affects only the particle yields and does not change the shape of the spectra.  
However, the vertex finding efficiency depends on the beam luminosity.  The number quoted in this 
row is the level of fluctuation in the vertex finding efficiency with beam luminosity, 4\%.

Conversion of our measurements to cross-sections must also account for an additional 
7.3\% uncertainty in the measured NSD trigger cross-section (26$\pm$1.9 mb) and for the 
86\% efficiency of the BBC trigger detectors.

\section{Results \label{Results}}

\subsection{Spectra \label{Spectra}}

\begin{figure*}[ht]
\epsfig{figure=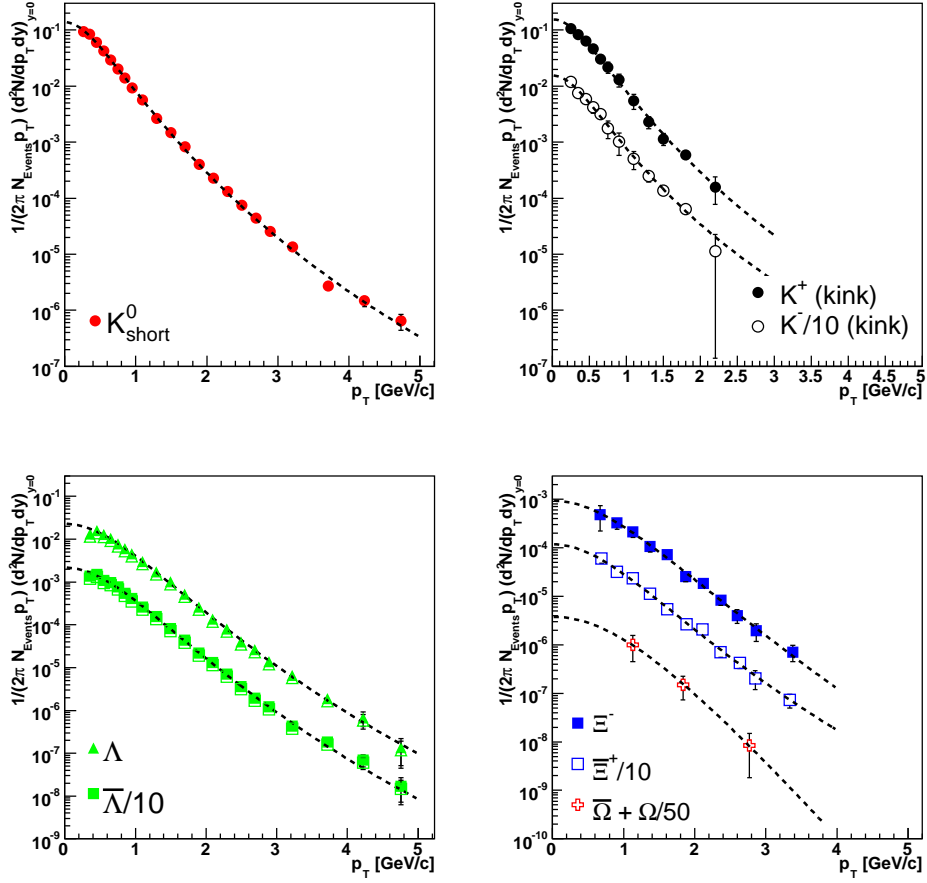,width=1.5\columnwidth} 
\caption{Corrected mid-rapidity ($|y|<0.5$) \pt~spectra for
\kp, \km, \Ks, $\Lambda$, $\Xi$, and $\Omega$.  $\Lambda$ spectra that have been 
corrected for feed-down are shown as open symbols in the $\Lambda$ panel.  The dashed lines 
are fits using Equation \ref{equ:Levy} except for the $\Omega+\overline{\Omega}$ where the 
fit uses Equation \ref{equ:mtexp}.  The error bars displayed include systematic errors while 
the fits were done using statistical errors only for all species except the charged kaons.}
\label{fig:spectra}
\end{figure*}

The fully corrected \pt~spectra for \kp, \km, \Ks, $\Lambda$, $\Xi$, and $\Omega$ 
are shown in Figure \ref{fig:spectra}.  The measured spectra cover only a 
limited range in transverse momentum and therefore an appropriately parameterized 
function is needed to extrapolate into the unmeasured \pt~regions 
for the yield determination.  In the past, exponential functions such 
as that given in Equation \ref{equ:mtexp}
have been used to extrapolate spectra from $p+p$ collisions to low 
transverse momentum while QCD-inspired power-law functions (see Equation 
\ref{equ:powerLaw}) seem to provide a better description of the high 
\pt~($\gtrsim$3 GeV/$c$) region \cite{UA5,UA5:87,Bocquet96,Hagedorn}.  
The \pt~coverage of the STAR detector for strange particles is large enough 
that a function which accounts for both the power-law component of the spectra 
and the low \pt~turnover becomes necessary to describe the data.  
A form that has been suggested is the L\'evy function given by Equation 
\ref{equ:Levy} \cite{Wilk}.  
 
\begin{equation}
\frac{1}{2\pi p_{T}}\frac{d^{2}N}{dy dp_{T}} =
Ae^{\frac{-m_{T}}{T}}
\label{equ:mtexp}
\end{equation}
\begin{equation}
\frac{1}{2\pi p_{T}}\frac{d^{2}N}{dy dp_{T}} =
B(1+\frac{p_{T}}{p_{0}})^{-n}
\label{equ:powerLaw}
\end{equation}
\begin{equation}
\begin{split}
\frac{1}{2\pi p_{T}}\frac{d^{2}N}{dy dp_{T}} =
\frac{dN}{dy}\frac{(n-1)(n-2)}{2\pi nC[nC+m_{0}(n-2)]} \\
\times\left(1+\frac{\sqrt{p_{T}^{2}+m_{0}^{2}}-m_{0}}{nC}\right)^{-n}
\label{equ:Levy}
\end{split}
\end{equation}
\begin{table}[h]
\begin{center}
\setlength\extrarowheight{2pt}
\begin{tabular}{|c|c|c|c|c|c|c|c|c|c|c|}
\hline                      & \multicolumn{2}{c}{$\mathbf{m_{T}{\bf -exponential}}$}\vline & 
                              \multicolumn{2}{c}{\textbf{Power-Law}}\vline & 
                              \multicolumn{2}{c}{\textbf{L\'evy}}\vline \tabularnewline
\hline \textbf{Particle}    & $\mathbf{\chi^{2}/ndf}$ & \textbf{ndf} & $\mathbf{\chi^{2}/ndf}$ & 
                              \textbf{ndf} & $\mathbf{\chi^{2}/ndf}$ & \textbf{ndf} \tabularnewline
\hline $\mathrm{K^{0}_{S}}$     & 15 & 22 & 1.5 & 21 & 0.89 & 19 \tabularnewline
\hline $\mathrm{K^{+}}$ (kinks) & 3.1 & 11 & 7.0 & 10 & 0.40 & 9 \tabularnewline
\hline $\mathrm{K^{-}}$ (kinks) & 9.4 & 11 & 5.0 & 10 & 0.30 & 9 \tabularnewline
\hline $\Lambda$                & 4.5 & 22 & 3.3 & 21 & 0.81 & 18 \tabularnewline
\hline $\overline{\Lambda}$     & 4.7 & 22 & 3.1 & 21 & 0.99 & 18 \tabularnewline
\hline $\Xi^{-}$                & 0.84 & 9 & 1.4 & 8 & 0.76 & 8 \tabularnewline
\hline $\overline{\Xi}^{+}$     & 1.4 & 9 & 0.96 & 8 & 0.83 & 8 \tabularnewline
\hline $\Omega^{-}+\overline{\Omega}^{+}$ & 0.13 & 1 & \multicolumn{4}{c}{--}\vline \tabularnewline
\hline
\end{tabular}
\caption{A summary of $\chi^2$ per degree of freedom values for different fit-functions
to the \pt~spectra.  The $\Omega^{-}+\overline{\Omega}^{+}$ statistics are considered insufficient 
for a fit comparison.}
\label{tab:chi2}
\end{center}
\end{table}
\begin{table}[ht]
\begin{center}
\setlength\extrarowheight{2pt}
\begin{tabular}{|c|c|c|c|c|c|c|c|c|}
\hline \textbf{Particle}    & 
       \textbf{dN/dy, $|y|<0.5$} & 
       \textbf{Stat. Err.} &
       \textbf{Sys. Err.} \tabularnewline
\hline $\mathrm{K^{0}_{S}}$      & 0.134 & 0.003 & 0.011 \tabularnewline
\hline $\mathrm{K^{+}}$ (kinks)  & 0.140 & 0.006 & 0.008 \tabularnewline
\hline $\mathrm{K^{-}}$ (kinks)  & 0.137 & 0.006 & 0.007  \tabularnewline
\hline $\Lambda$                 & 0.0436 & 0.0008 & 0.0040 \tabularnewline
\hline $\overline{\Lambda}$      & 0.0398 & 0.0008 & 0.0037 \tabularnewline
\hline $\Lambda$ (FD)            & 0.0385 & 0.0007 & 0.0035 \tabularnewline
\hline $\overline{\Lambda}$ (FD) & 0.0351 & 0.0007 & 0.0032 \tabularnewline
\hline $\Xi^{-}$                 & 0.0026 & 0.0002 & 0.0009 \tabularnewline
\hline $\overline{\Xi}^{+}$      & 0.0029 & 0.0003 & 0.0010 \tabularnewline
\hline $\Omega^{-}+\overline{\Omega}^{+}$ & 0.00034 & 0.00016 & 0.0001 \tabularnewline
\hline
\end{tabular}
\caption{A summary of mid-rapidity NSD yields for
measured strange particles.  Numbers in rows marked (FD) 
have been corrected for feed-down as described in section 
\ref{Feeddown}.}
\label{tab:yields}
\end{center}
\end{table}
\begin{table}[!ht]
\begin{center}
\setlength\extrarowheight{2pt}
\begin{tabular}{|c|c|c|c|c|c|c|c|c|}
\hline \textbf{Particle}    & 
       $\mathbf{\langle{\bf p_{T}}\rangle}$\textbf{(GeV/$\mathbf{c}$)} &
       \textbf{Stat. Err.} &
       \textbf{Sys. Err.} \tabularnewline
\hline $\mathrm{K^{0}_{S}}$      & 0.605 & 0.010 & 0.023 \tabularnewline
\hline $\mathrm{K^{+}}$ (kinks)  & 0.592 & 0.071 & 0.014 \tabularnewline
\hline $\mathrm{K^{-}}$ (kinks)  & 0.605 & 0.072 & 0.014 \tabularnewline
\hline $\Lambda$                 & 0.775 & 0.014 & 0.038 \tabularnewline
\hline $\overline{\Lambda}$      & 0.763 & 0.014 & 0.037 \tabularnewline
\hline $\Lambda$ (FD)            & 0.762 & 0.013 & 0.037 \tabularnewline
\hline $\overline{\Lambda}$ (FD) & 0.750 & 0.013 & 0.037 \tabularnewline
\hline $\Xi^{-}$                 & 0.924 & 0.120 & 0.053 \tabularnewline
\hline $\overline{\Xi}^{+}$      & 0.881 & 0.120 & 0.050 \tabularnewline
\hline $\Omega^{-}+\overline{\Omega}^{+}$ & 1.08 & 0.29 & 0.09 \tabularnewline 
\hline
\end{tabular}
\caption{A summary of mid-rapidity \mpt~for measured strange particles.  Feeddown 
corrected numbers for $\Lambda$ and $\overline{\Lambda}$ are the same as 
the non-feed-down corrected values within statistical errors.}
\label{tab:meanpt}
\end{center}
\end{table}
\begin{table*}[ht]
\begin{center}
\setlength\extrarowheight{2pt}
\begin{tabular}{|c|c|c|c|c|c|} 
\hline \textbf{Particle} & 
       \textbf{STAR dN/dy ($|\textbf{y}|$ $<$ 0.5)} & 
       \textbf{UA5 Yield} & 
       \textbf{STAR Yield (scaled to UA5 y)}  \tabularnewline 
\hline \Ks~& 0.134 $\pm$ 0.011 &
             0.73  $\pm$ 0.18, $|\textbf{y}| < 3.5$& 
             0.626 $\pm$ 0.051 \tabularnewline 
\hline $\Lambda + \overline{\Lambda}$ & 
       0.0834 $\pm$ 0.0056 & 
       N/A & 
       0.272 $\pm$ 0.018 \tabularnewline 
\hline $\Lambda + \overline{\Lambda}$ (FD) & 
       0.0736 $\pm$ 0.0048 &
       0.27 $\pm$ 0.09, $|\textbf{y}| < 2.0$ & 
       0.240 $\pm$ 0.016 \tabularnewline 
\hline $\Xi + \overline{\Xi}$ & 
       0.0055 $\pm$ 0.0014 & 
       $0.03{^{+0.04}_{-0.02}}$, $|\textbf{y}| < 3.0$ & 
       0.0223 $\pm$ 0.0057 \tabularnewline 
\hline
\end{tabular}
\caption{A comparison of yields from UA5 (\Ks~from \cite{UA5:87}, $\Lambda$ from \cite{UA5}) 
and NSD yields from STAR.  The STAR entries in the last column have been scaled to the UA5 
acceptance using \textsc{Pythia} \cite{pythia}.  The STAR errors include systematics.  The 
UA5 errors shown include their estimated 20\% systematic error.}
\label{tab:UA5dndy}
\end{center}
\end{table*}

\begin{table}[ht]
\begin{center}
\setlength\extrarowheight{2pt}
\begin{tabular}{|c|c|c|c|} 
\hline \textbf{Particle} & 
\textbf{STAR $\langle \textbf{p}_{\textbf{T}} \rangle ~(|\textbf{y}|< 0.5)$} & 
       \textbf{UA5 $\Lambda~\langle \textbf{p}_{\textbf{T}} \rangle$} \tabularnewline 
\hline
        \Ks~ & 0.61 $\pm$ 0.02 & 
       $0.53{^{+0.13}_{-0.12}}$, $|\textbf{y}|<2.5$ \tabularnewline 
\hline
       $\Lambda + \overline{\Lambda}$ & 
       0.77 $\pm$ 0.04 & 
       $0.8{^{+0.26}_{-0.21}}$, $|\textbf{y}|<2.0$\tabularnewline 
\hline
       $\Xi + \overline{\Xi}$ & 
       0.903 $\pm$ 0.13 & 
       $0.8{^{+0.4}_{-0.2}}$, $|\textbf{y}|<3.0$\tabularnewline 
\hline
\end{tabular}
\caption{A comparison of \mpt~[GeV/$c$] from UA5 and STAR.  STAR errors include systematics.  The 
UA5 errors shown include their estimated 20\% systematic error.}
\label{tab:UA5mpt}
\end{center}
\end{table}
where $A$, $T$, $B$, $p_{0}$, $n$, $\frac{dN}{dy}$, $C$, and $m_{0}$ are fit parameters.  
Attempts were made to fit the \pt~spectra for our measured species with 
all three forms.  A summary of the resulting $\chi^2/$ndf from each fit 
is given in Table \ref{tab:chi2} for each of the measured species.  
The mid-rapidity yields and mean transverse momenta quoted below were 
determined from the best fitting form which, for all species, was the 
L\'evy form (Equation \ref{equ:Levy}).  The measured mid-rapidity yields 
and feed-down corrected yields are presented in Table \ref{tab:yields}.  
The measured mean transverse momenta are presented in Table \ref{tab:meanpt}.

Initially, we compare our measurement of neutral strange particles
to similar experiments at this energy.  The closest comparison can
be made to the \sps~(Super Proton-Antiproton Synchrotron)
experiments of UA1-UA5 using the \pbarp~beam.  Only UA5 published strange 
particle measurements at \sqs=200 GeV \cite{UA5,UA5:87}, with others at \sqs=546 GeV 
\cite{UA5:546} and 900 GeV \cite{UA5,UA5:87} while UA1 published high statistics strange particle
measurements at \sqs=630 GeV (\cite{Bocquet96} and references cited therein). 

It is worth noting that the UA5 $\Lambda$ sample consisted of only 
168 ``manually sorted" candidates \cite{UA5}, whereas the STAR sample
consists of fifty-eight thousand candidates.

Table \ref{tab:UA5dndy} compares the values of \dndy~and obtained from the 
STAR \pt~spectra to the published values from the UA5 experiment at
\sps~\cite{UA5:87} measured with a larger rapidity interval.  In the 
last column, the STAR data is scaled by a factor, obtained 
via \textsc{Pythia} \cite{pythia} simulation, to account for the difference in rapidity 
coverage of the two experiments.  UA5 measured \Ks~with
\midy 2.5, $\Lambda$ with \midy 2.0, and $\Xi$ with \midy 3.0.  STAR 
measures only in the region \midy 0.5.  The STAR scaled yields are found to be in 
agreement with the measurement from UA5 and have greatly improved on the precision.

Table \ref{tab:UA5mpt} compares the \mpt~of the two experiments. 
It was verified, using \textsc{Pythia}, that the dependence of \mpt~on the different 
rapidity intervals between STAR and UA5 is small, \textit{i.e.}  2-3\%. 
Therefore, the STAR \mpt~measurement is compared to UA5 without further 
scaling and is found to have improved on the precision.

\subsection{Transverse Mass Scaling \label{mt-scalingSection}}
\begin{figure*}[ht]
\subfigure[~Transverse mass mid-rapidity ($|y|<0.5$) spectra for $\pi$, $\text{K}^{+}$, 
\Ks, p, $\Lambda$, $\Xi$.]{
\label{fig:mtNOscaling}
\includegraphics[width=0.7\columnwidth]{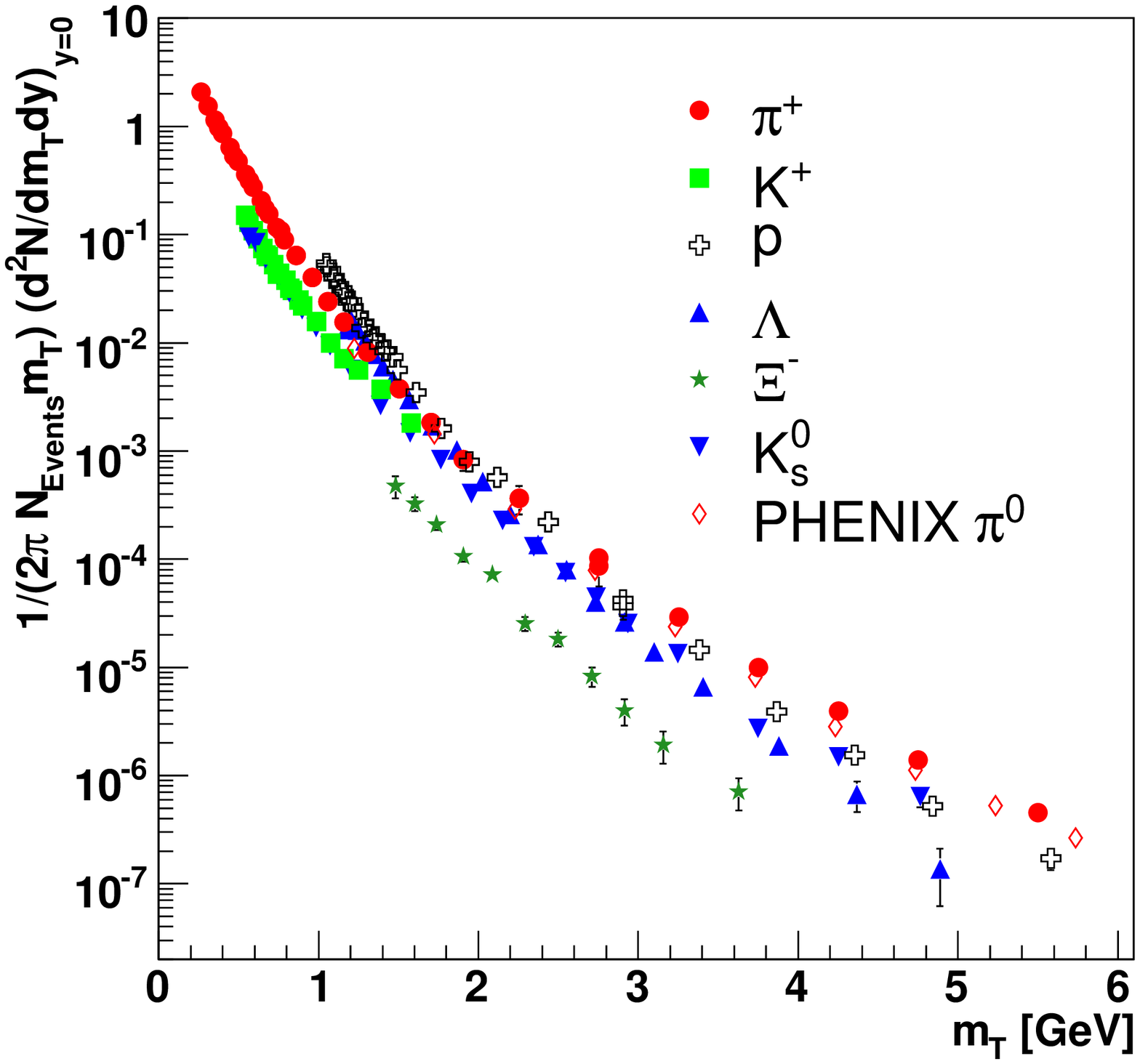}}
\subfigure[~Scaled transverse mass mid-rapidity ($|y|<0.5$) spectra for $\pi^{+}$, $\text{K}^{+}$, 
\Ks, p, $\Lambda$, $\Xi$.]{
\label{fig:mtscaling}
\includegraphics[width=0.7\columnwidth]{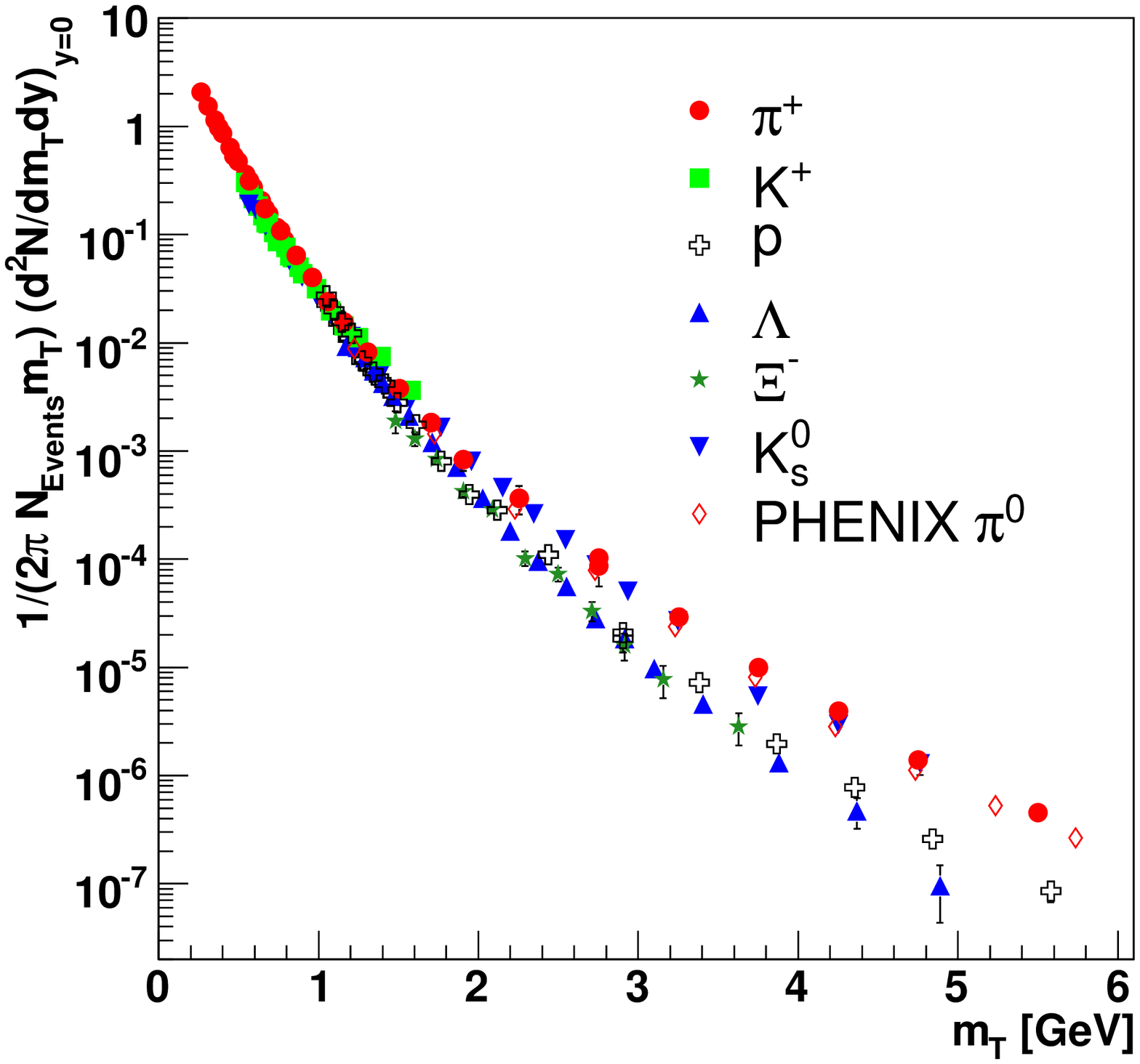}}
\subfigure[~Ratio of data to meson power-law fit for each data point in Figure \ref{fig:mtscaling}.]{
\label{fig:mesonFit}
\includegraphics[width=0.7\columnwidth]{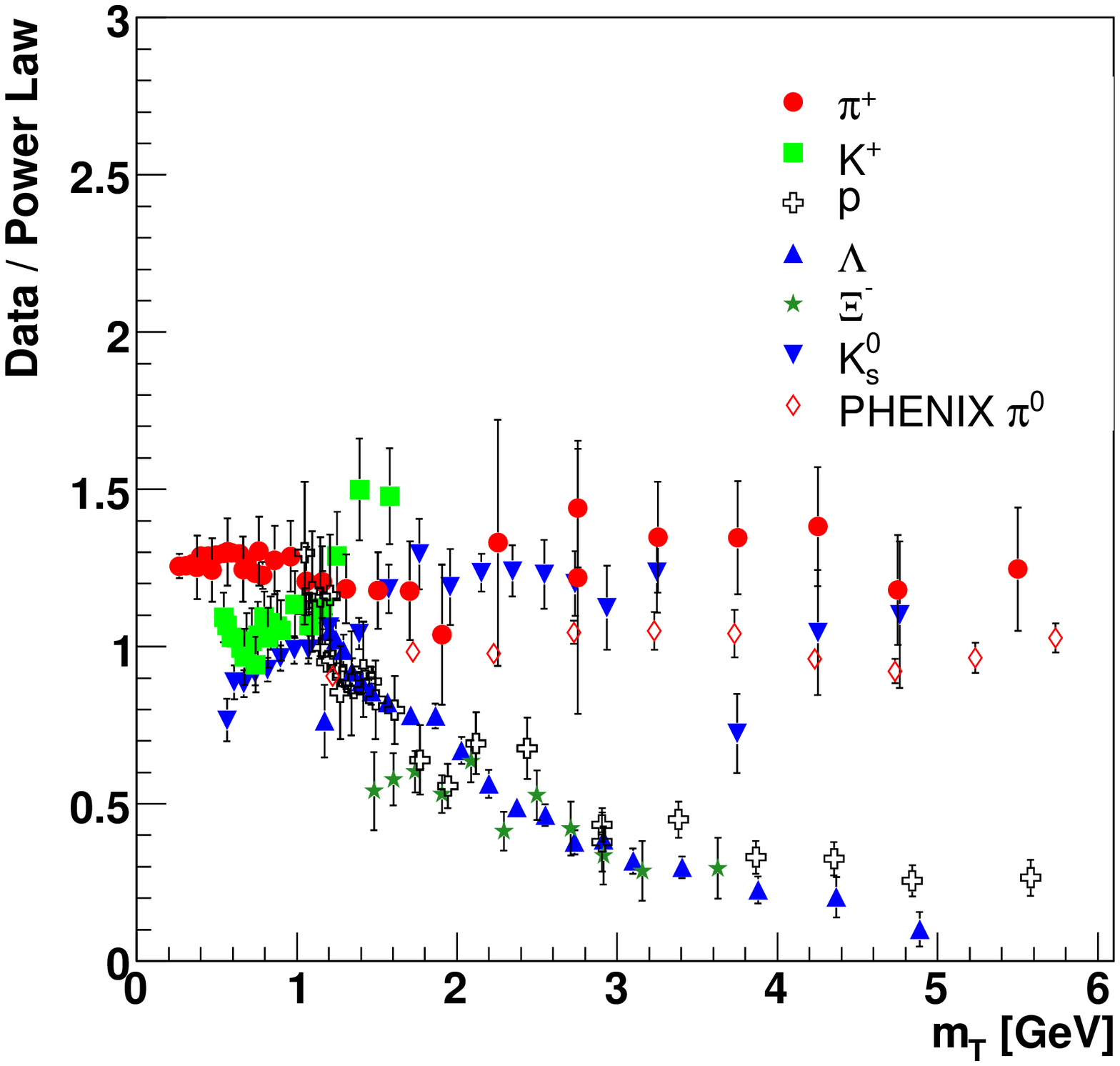}}
\subfigure[~Ratio of data to baryon power-law fit for each data point in Figure \ref{fig:mtscaling}.]{
\label{fig:baryonFit}
\includegraphics[width=0.7\columnwidth]{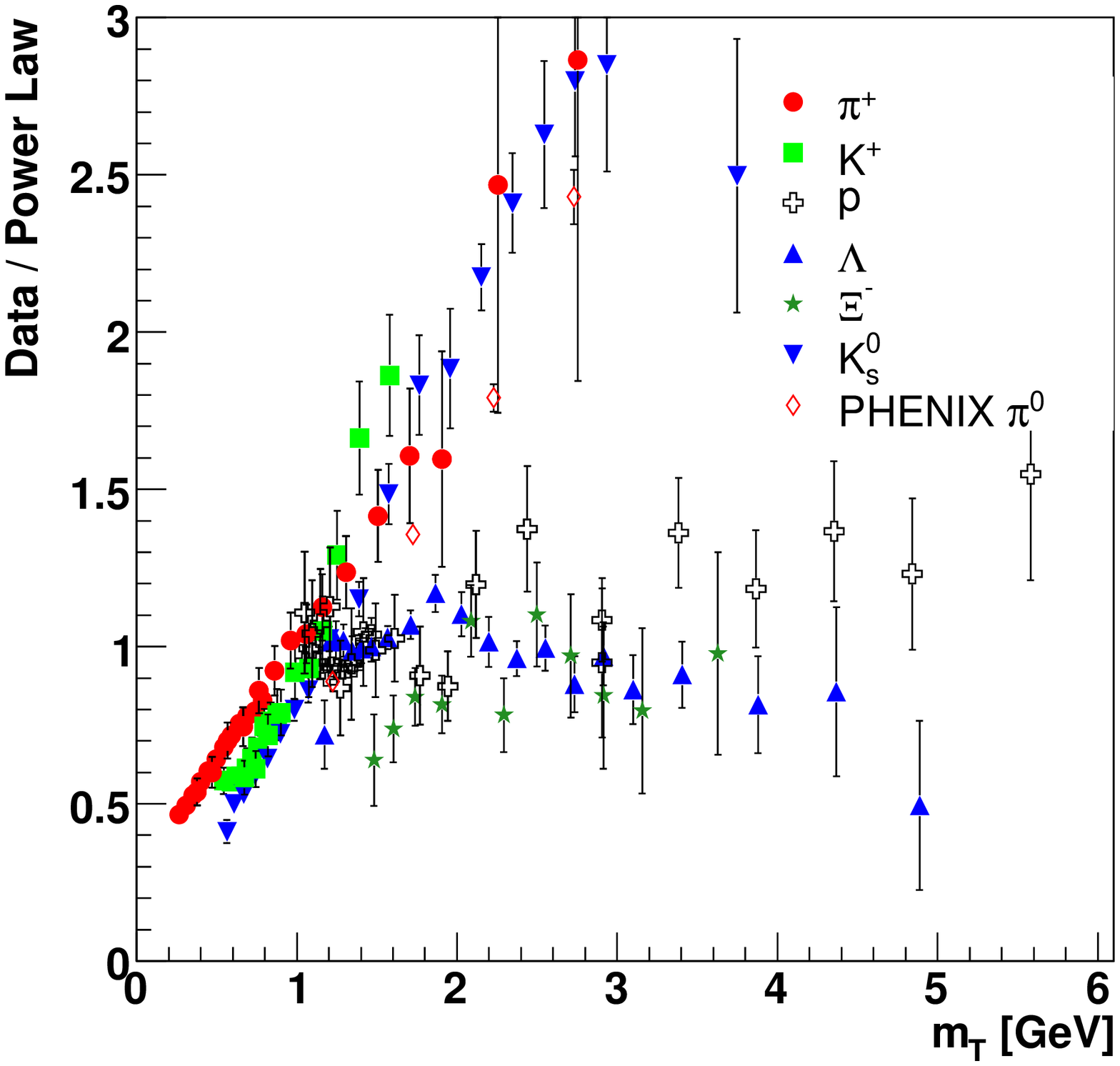}}
\caption{Comparisons of scaled and un-scaled transverse mass spectra from $p+p$ collisions 
in STAR and PHENIX at $\sqrt{\mathrm{s}}$=200 GeV.  $\pi$, K, and p spectra are from 
\cite{STAR-pi-k-p,STAR-TOF,STAR-rDEDX} while the PHENIX $\pi^{0}$ spectrum is from \cite{phenix-pi0}.  
Error bars include systematics.}
\label{mtscalingFigures}
\end{figure*}
It has been noted previously that the identified particle spectra from $p+p$ 
collisions at ISR energies \cite{ISR75,ISR81} seem to sample an approximately 
universal curve when plotted versus transverse mass \cite{Wong}, an effect 
termed ``\mt-scaling".  More recently, data from heavy-ion collisions at RHIC have been 
shown to scale in transverse mass over the measured range available \cite{Schaffner-Bielich02}.  
Transverse mass spectra from identified hadrons at $\sqrt{\mathrm{s}}$=540 GeV and 630 GeV 
$p+\overline{p}$ collisions at Sp$\overline{\mathrm{p}}$S have also been shown to exhibit the 
same behavior up to at least 2.5 GeV \cite{Schaffner-Bielich02}.  The degree to which \mt-scaling 
is applicable and the resulting scaling factors have been used to argue for the presence of a gluon-saturated 
state (color-glass condensate) in heavy-ion collisions at RHIC energies \cite{Schaffner-BielichCGC02}, 
though no such interpretation is applied to $p+p$ or $p+\overline{p}$ collisions.  Little discussion of 
the similarity of the results between $p+p$ and A+A has been provided.  
In Figure \ref{fig:mtNOscaling} we present the \Ks, $\Lambda$, 
and $\Xi$ \mt~spectra together with their antiparticles and with \mt~spectra 
for $\pi$, K, and p from previously published STAR $p+p$ results at $\sqrt{\mathrm{s}}$=200 GeV 
\cite{STAR-pi-k-p,STAR-TOF,STAR-rDEDX}.  The PHENIX $\pi^{0}$ spectrum from $p+p$ collisions at the same 
energy is also shown \cite{phenix-pi0}.

It is clear from Figure \ref{fig:mtNOscaling} that while the spectra appear to have 
qualitatively similar shapes, the yields are quite different.  Nevertheless, the shape similarities 
encourage us to find a set of scaling factors that would bring the 
spectra onto a single curve.  Figure \ref{fig:mtscaling} shows the result of scaling 
with the set of factors shown in Table \ref{tab:mtScalingFactors}.  These factors were 
chosen so as to match the $\pi$, K, and p spectra at an \mt~of 1 GeV.  The higher mass spectra 
are then scaled to match the $\pi$, K, and p spectra in their respective regions of overlap.

While the low-\mt~region seems to show reasonable agreement between all the measured species the region 
above \mt~$\sim$2 GeV shows an interesting new effect.  The meson spectra appear to be 
harder than the baryon spectra with as much as an order of magnitude difference 
developing by 4.5 GeV in \mt.  In order to quantify the degree of agreement, a power-law 
function was fit to all the scaled meson and baryon \mt~spectra separately.  The 
ratio of data to each fit was taken for each point in Figure \ref{fig:mtscaling}.  
The data-to-fit ratio is shown for the meson fit in Figure \ref{fig:mesonFit} and 
the baryon fit in Figure \ref{fig:baryonFit}.

This is the first time such a meson-baryon effect has been noticed in $p+p$ 
collisions.  This effect is observable due to the high \pt~(and therefore high-\mt) 
coverage of the strange particle and relativistic rise spectra \cite{STAR-rDEDX}.  
The harder meson spectrum in the jet-like high-\mt~region may indicate that for a 
given jet energy, mesons are produced with 
higher transverse momentum than baryons.  This effect would be a simple reflection 
of the fact that meson production from fragmentation requires only a (quark,anti-quark) 
pair while baryon production requires a (di-quark,anti-di-quark) pair.  The 
difference between the baryon and meson curves appears to be increasing over our 
measured range, and it will be interesting to see, with greater statistics, what level 
of separation is achieved and whether or not the spectra eventually become parallel.
\begin{table}[h]
\begin{center}
\setlength\extrarowheight{2pt}
\begin{tabular}{|c|c|c|c|c|c|c|}
\hline  & 
	$\mathbf{\pi}$ & 
	$\textbf{K}$ &
	\textbf{p} &
	$\mathbf{\Lambda}$ &
	$\mathbf{\Xi}$ \tabularnewline
\hline \textbf{Scaling Factor}  &
	1.0 &  
	2.0 &  
	0.6\footnote{Data from \cite{STAR-pi-k-p} were scaled by 0.45} &  
	0.7 &  
	4.0 \tabularnewline 
\hline \textbf{Scaled at $\mathrm{m_{T}}$}  &
	1.0 &  
	1.0 &  
	1.0 &  
	1.5 &  
	1.5 \tabularnewline 
\hline
\end{tabular}
\caption{A summary of scaling factors applied to the transverse mass 
spectra in Figure \ref{fig:mtscaling}.  The second row lists the transverse 
mass (in GeV) at which a given particle is scaled to match the other spectra.}
\label{tab:mtScalingFactors}
\end{center}
\end{table}
\begin{figure*}
\subfigure[~Anti-particle to particle ratios vs strangeness content.]{
\label{fig:allRatios}
\includegraphics[width=0.67\columnwidth]{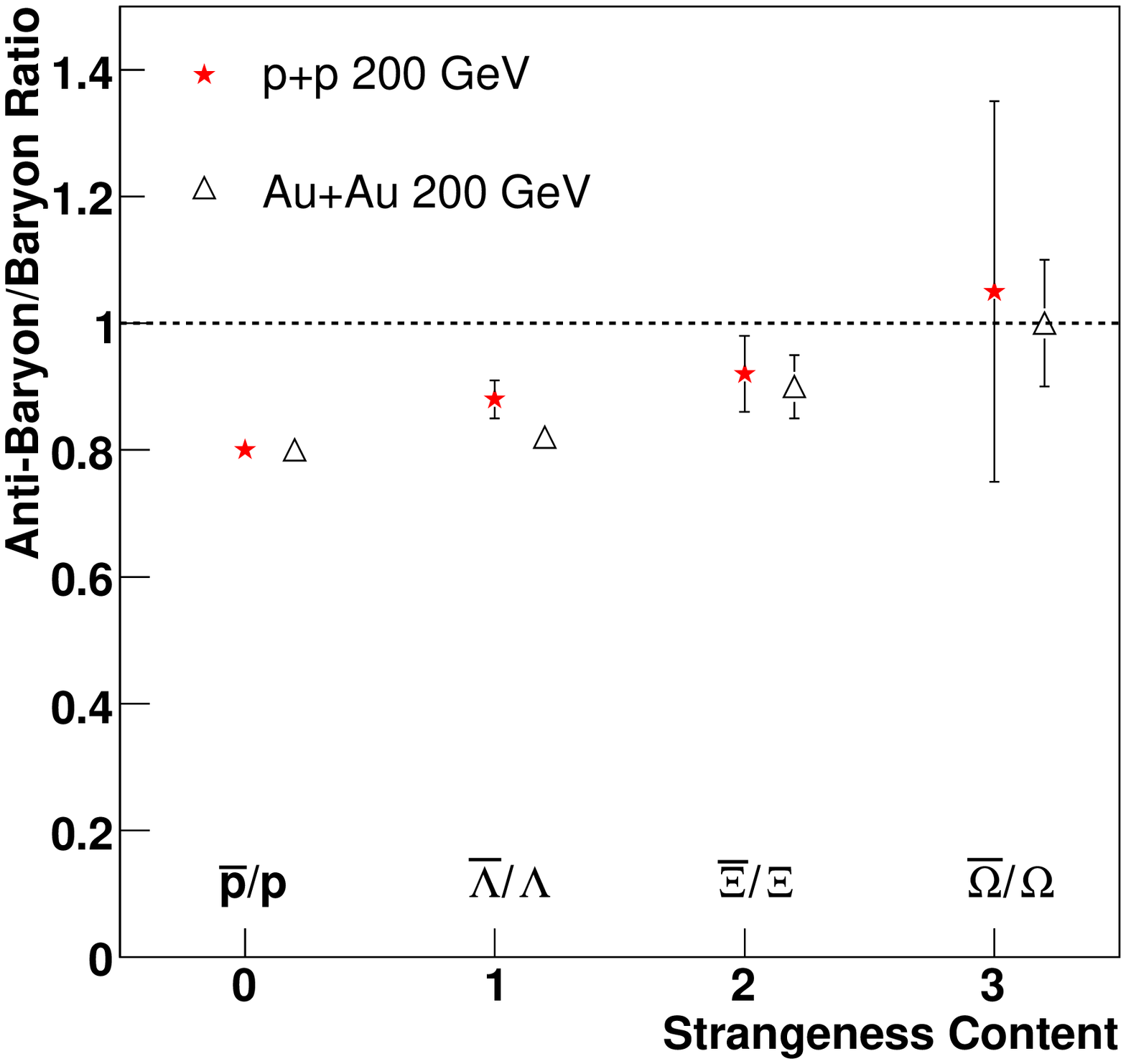}}
\subfigure[~$\overline{\Lambda}$ to $\Lambda$ vs \pt.]{
\label{fig:ratioLam}
\includegraphics[width=0.67\columnwidth]{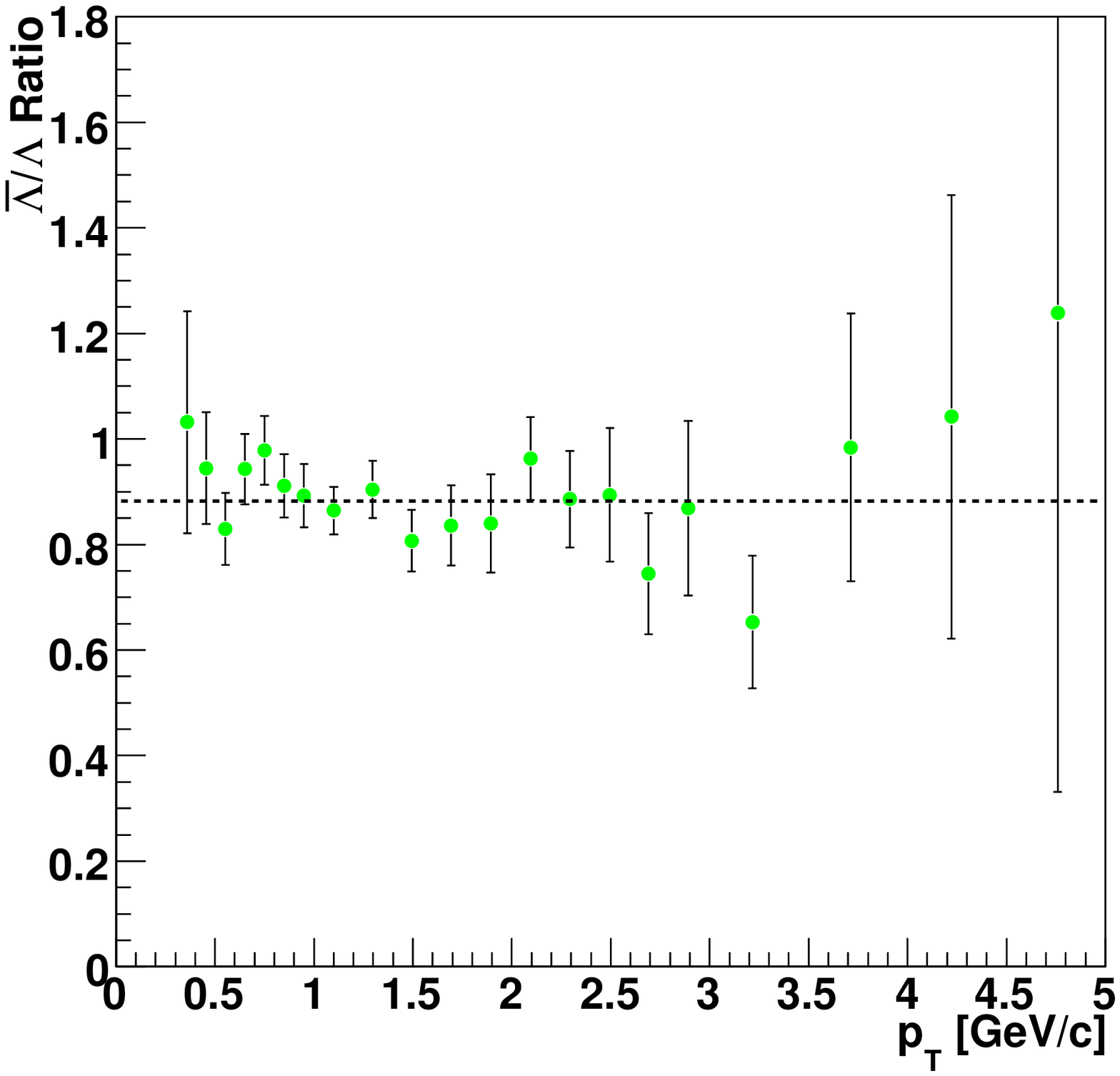}}
\subfigure[~$\overline{\Xi}^{+}$ to $\Xi^{-}$ vs \pt.]{
\label{fig:ratioXi}
\includegraphics[width=0.67\columnwidth]{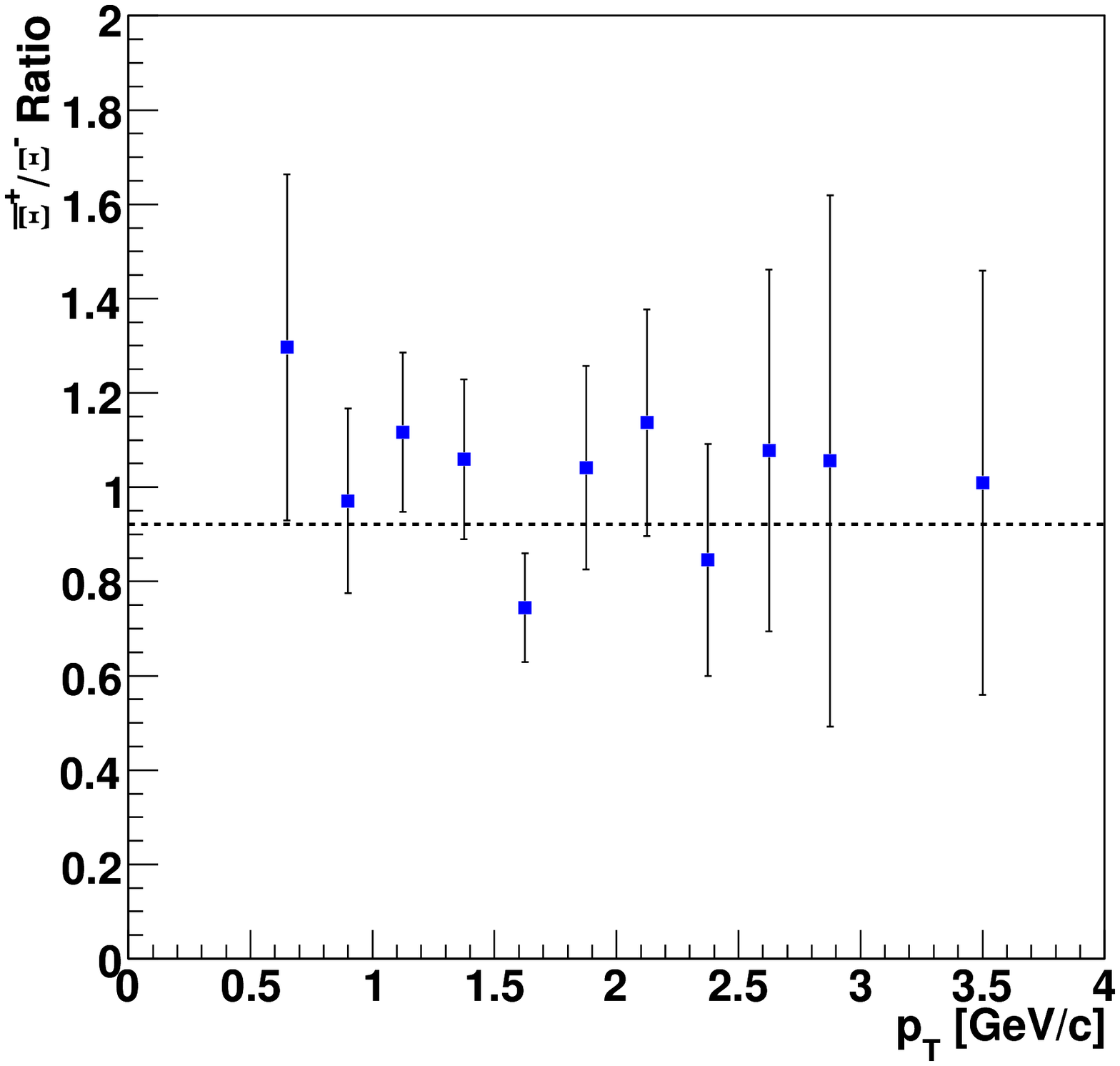}}
\caption{Mid-rapidity ($|y|<0.5$) ratios of $\overline{\Lambda}$ to $\Lambda$ and 
$\overline{\Xi}^{+}$ to $\Xi^{-}$ vs \pt.  The dashed lines in \ref{fig:ratioLam} and 
\ref{fig:ratioXi} are the error-weighted means over the measured \pt~range, 0.882$\pm$0.017 for 
$\overline{\Lambda}/\Lambda$, 0.921$\pm$0.062 for $\overline{\Xi}/\Xi$.  Figure 
\ref{fig:allRatios} shows the \pt~averaged ratio for our measured species compared with 
measurements from Au+Au.  The error bars are statistical only.  The dashed line 
in \ref{fig:allRatios} is at unity for reference.}
\label{fig:ratioFigures}
\end{figure*}

\subsection{Particle Ratios \label{Ratios}}

Figure \ref{fig:allRatios} shows the mean anti-baryon/baryon ratios 
($\overline{\text{B}}/\text{B}$) as a function of strangeness content 
for $p+p$ and Au+Au at $\sqrt{s} = 200$ GeV \cite{HarrisSQM03}.   
The ratios rise slightly with increasing strangeness 
content and are consistent within errors with those from Au+Au collisions at the same 
center-of-mass energy.  Although the $\overline{\text{B}}/\text{B}$ ratios are not 
unity for the protons and $\Lambda$ baryons, the deviation from unity may be explained 
by different parton distributions for the light quarks \cite{HERA}.  This may be 
sufficient to explain the observed deviation from unity without having to invoke 
baryon number transport over five units of rapidity.  
 
In the case of a quark jet, it is expected that there is a leading baryon as 
opposed to anti-baryon while there is no such distinction for a gluon jet.  
Therefore, making the assumption that at high \pt~the observed hadron production 
mechanisms are dominated by jet fragmentation, it is reasonable to 
expect that the $\overline{\text{B}}/\text{B}$ ratio will drop with increasing \pt.  
This has been predicted previously for calculations starting from as low as 2 GeV/$c$ 
\cite{Xin-Nian}.  Figures \ref{fig:ratioLam} and \ref{fig:ratioXi} 
show the $\overline{\Lambda}/\Lambda$ and $\overline{\Xi}/\Xi$ ratios as 
a function of transverse momentum respectively.   Although the errors shown in 
these figures are large, the ratios show no sign of decrease in the measured range.  The dotted 
horizontal line in each figure is the error-weighted average over the measured \pt~range.

One conclusion that could be drawn from the ratios in Figure \ref{fig:ratioFigures} is 
that particle production is not predominantly the result of quark-jet fragmentation 
over our measured range of \pt.

\subsection{Mean Transverse Momentum \label{meanPtSec}}

\begin{figure}[!ht]
\epsfig{figure=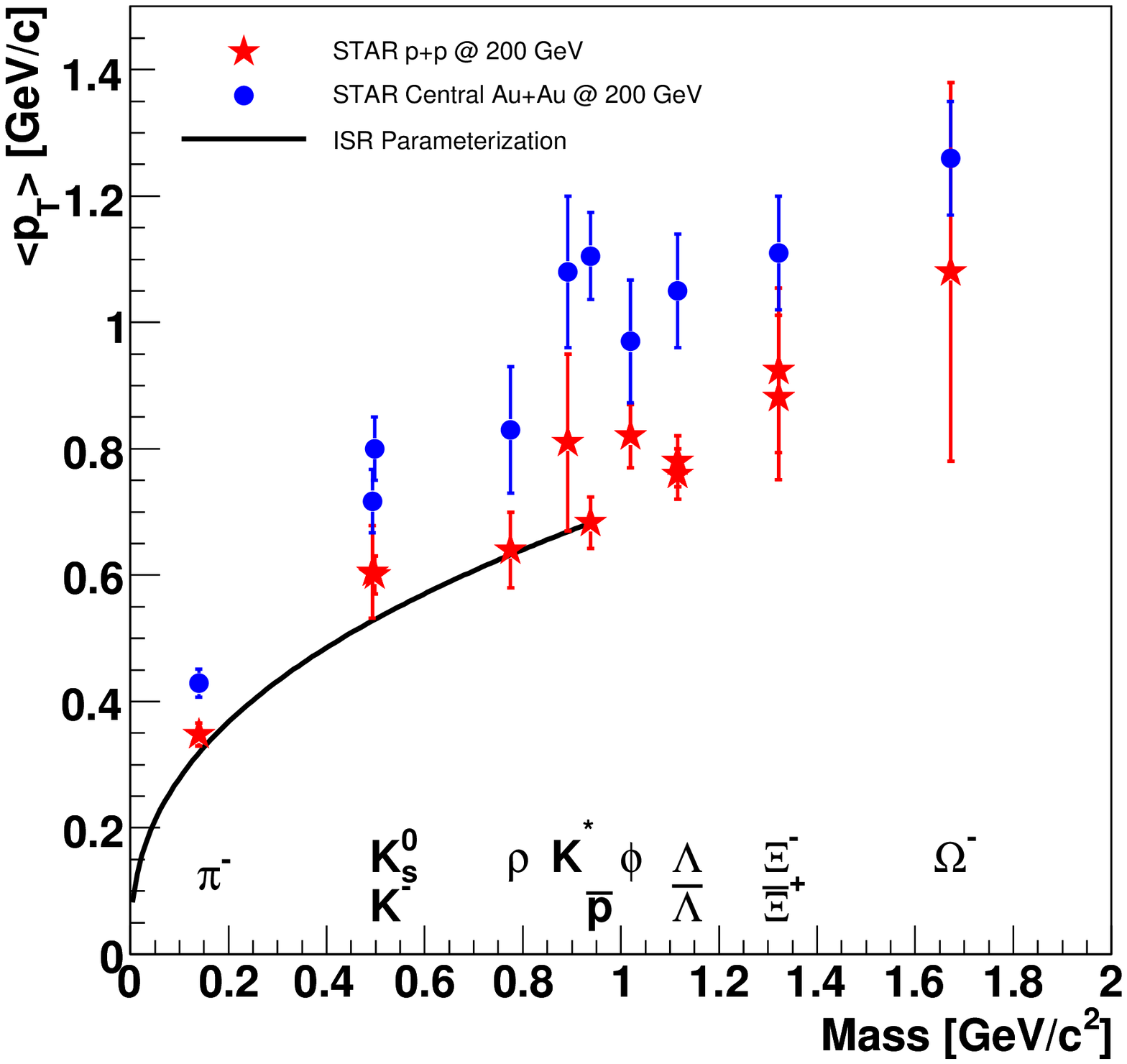,width=\columnwidth} 
\caption{\mpt~vs particle mass for different 
particles measured by STAR.  Error bars include systematic errors.  The ISR 
parameterization is given in reference \cite{ISRcurve}}.
\label{fig:mptmass}
\end{figure}

One means of partially characterizing the \pt~spectra from $p+p$ collisions 
is through the determination and comparison of the mean transverse momentum.  
In Figure \ref{fig:mptmass}, the \mpt~is shown for all particle species 
measured in both $p+p$ and central Au+Au collisions in STAR.  

In total, twelve particles in both systems are presented, covering a mass range 
of approximately 1.5 GeV/$c^2$.  The solid line is an empirical curve proposed 
originally \cite{ISRcurve} to describe the ISR \cite{ISR25GeV} and FNAL \cite{FNAL} 
data for $\pi$, K , and p only, at $\sqrt{s}$ = 25 GeV.  It is interesting that it 
fits the STAR lower mass particles from $p+p$ at $\sqrt{s}$ = 200 GeV remarkably well 
considering there is nearly an order of magnitude difference in collision energy.  
However, it is clear that a different parameterization would be needed to describe 
all of the STAR $p+p$ data.  The dependence of the inverse slope parameter, T 
(and therefore of the \mpt), on particle mass has previously been proposed to be due to an 
increasing contribution to the transverse momentum spectra from mini-jet production in 
$p+p$ and $p+\overline{p}$ collisions \cite{Dumitru}.  The contribution is 
expected to be even greater for higher mass particles \cite{Gyu92_2}.

\begin{figure}[ht]
\epsfig{figure=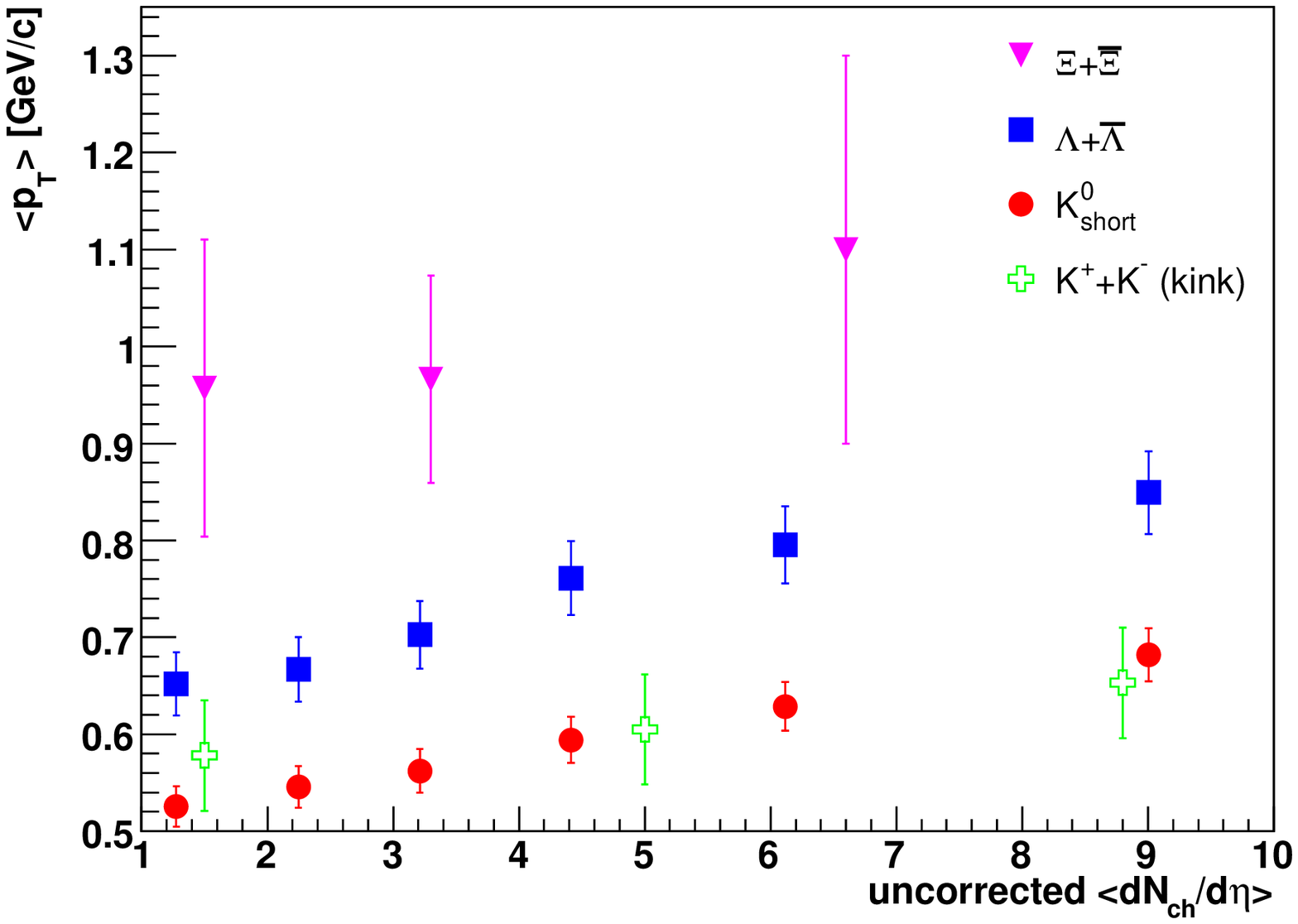,width=\columnwidth} 
\caption{\mpt~vs charged multiplicity for \kp, \km, \Ks, $\Lambda+\overline{\Lambda}$, 
and $\Xi+\overline{\Xi}$.  The points for $\Xi+\overline{\Xi}$ have been determined using 
only the measured region.  The error bars are statistical only.  See text for more details.}
\label{fig:mptmult}
\end{figure}

The available statistics allow a detailed study to be made.  
The mid-rapidity \pt~spectra can be binned according to eventwise charged particle 
multiplicity (uncorrected $dN_{\mathrm ch}/d\eta$) and the \mpt~determined in each bin.  We 
present in Figure \ref{fig:mptmult} the dependence of \mpt~on uncorrected 
charged particle multiplicity for \kp, \km, \Ks, $\Lambda$, and $\Xi$.  
\begin{figure*}
\subfigure[~R$_{pp}$ for \Ks.]{
\label{fig:RppK0}
\includegraphics[width=0.67\columnwidth]{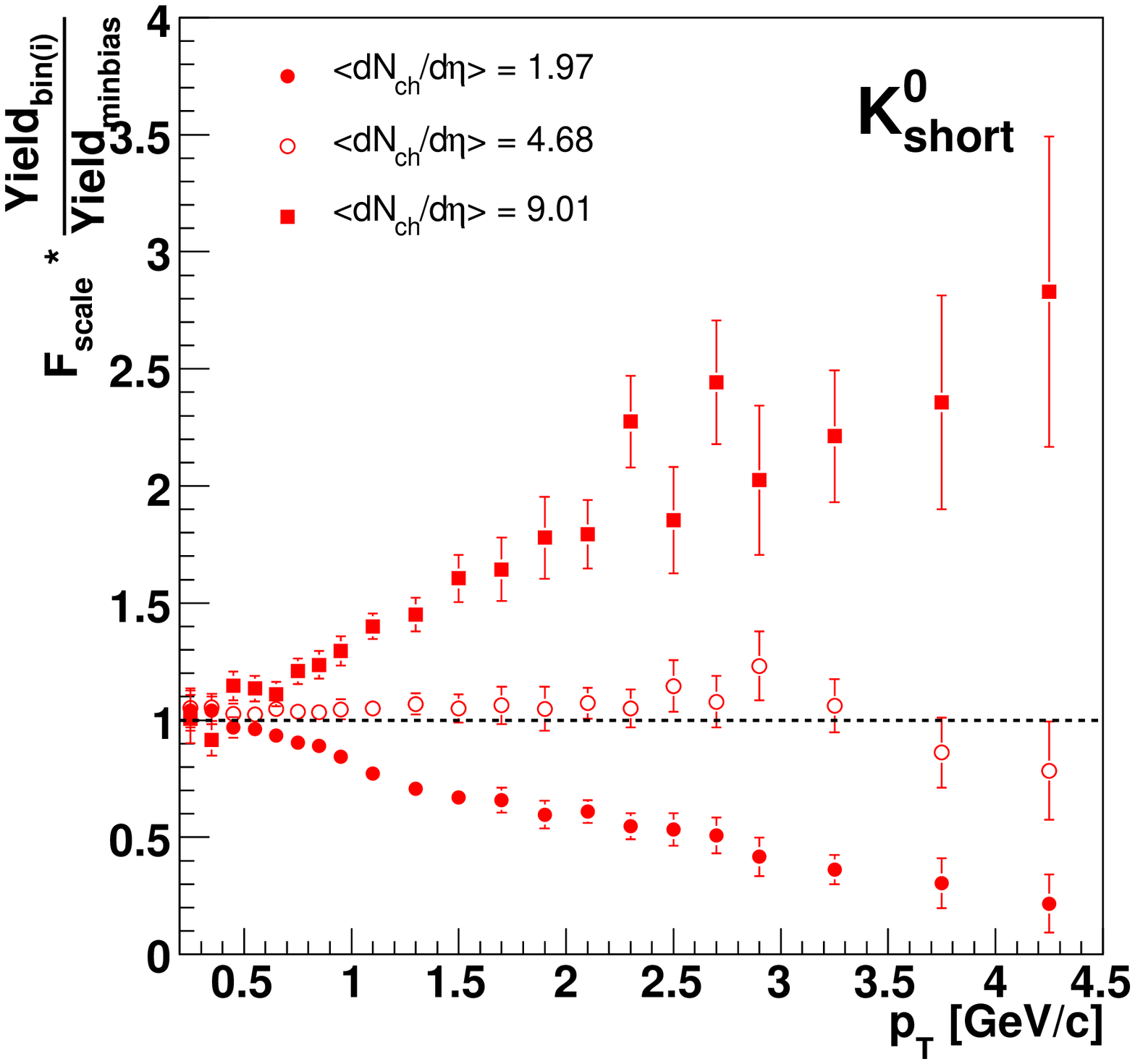}} 
\subfigure[~R$_{pp}$ for $\Lambda$.]{
\label{fig:RppLam}
\includegraphics[width=0.67\columnwidth]{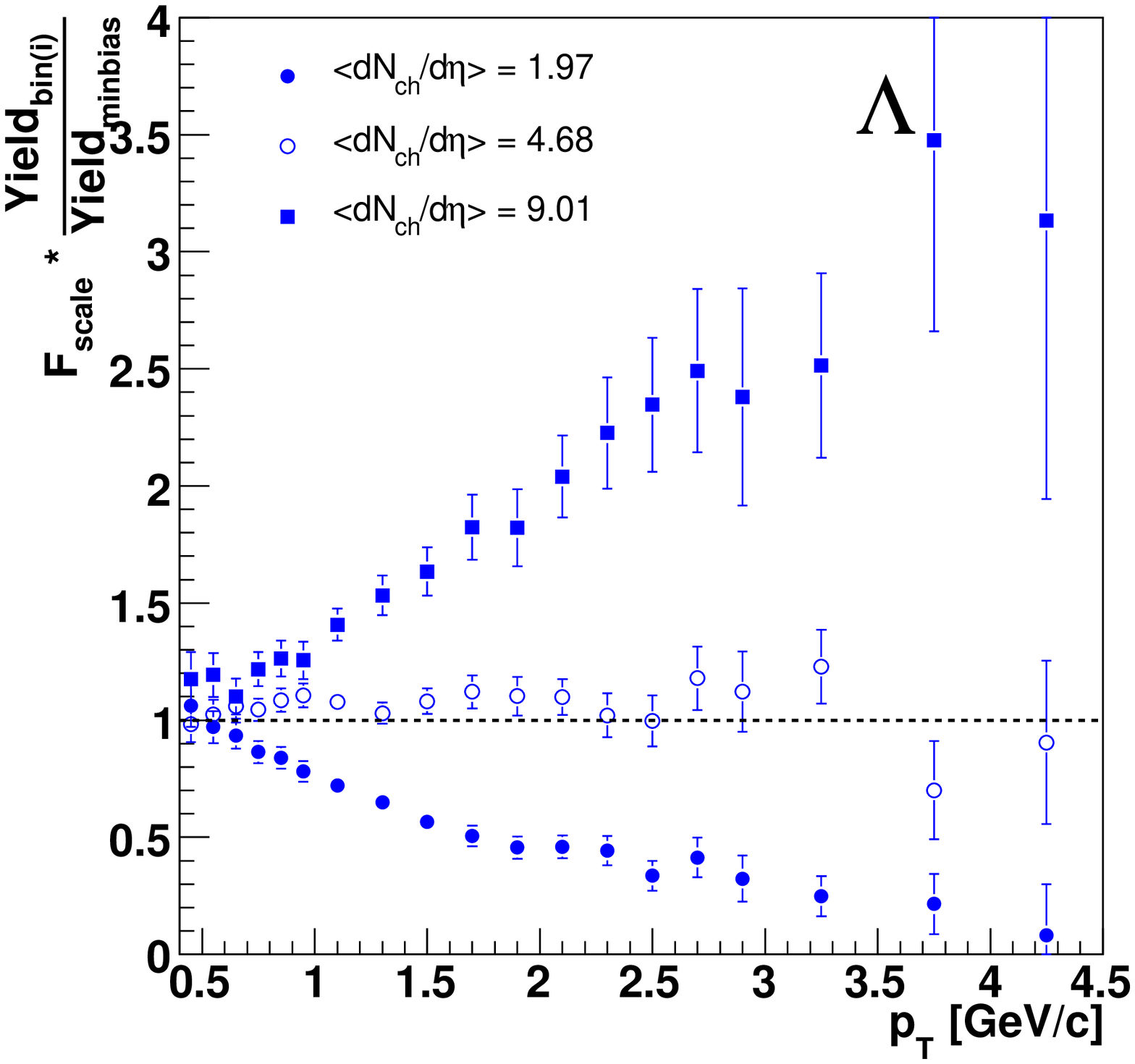}}
\subfigure[~Ratio of $\Lambda$ spectra to \Ks~spectra.]{
\label{fig:LamK0Ratio}
\includegraphics[width=0.67\columnwidth]{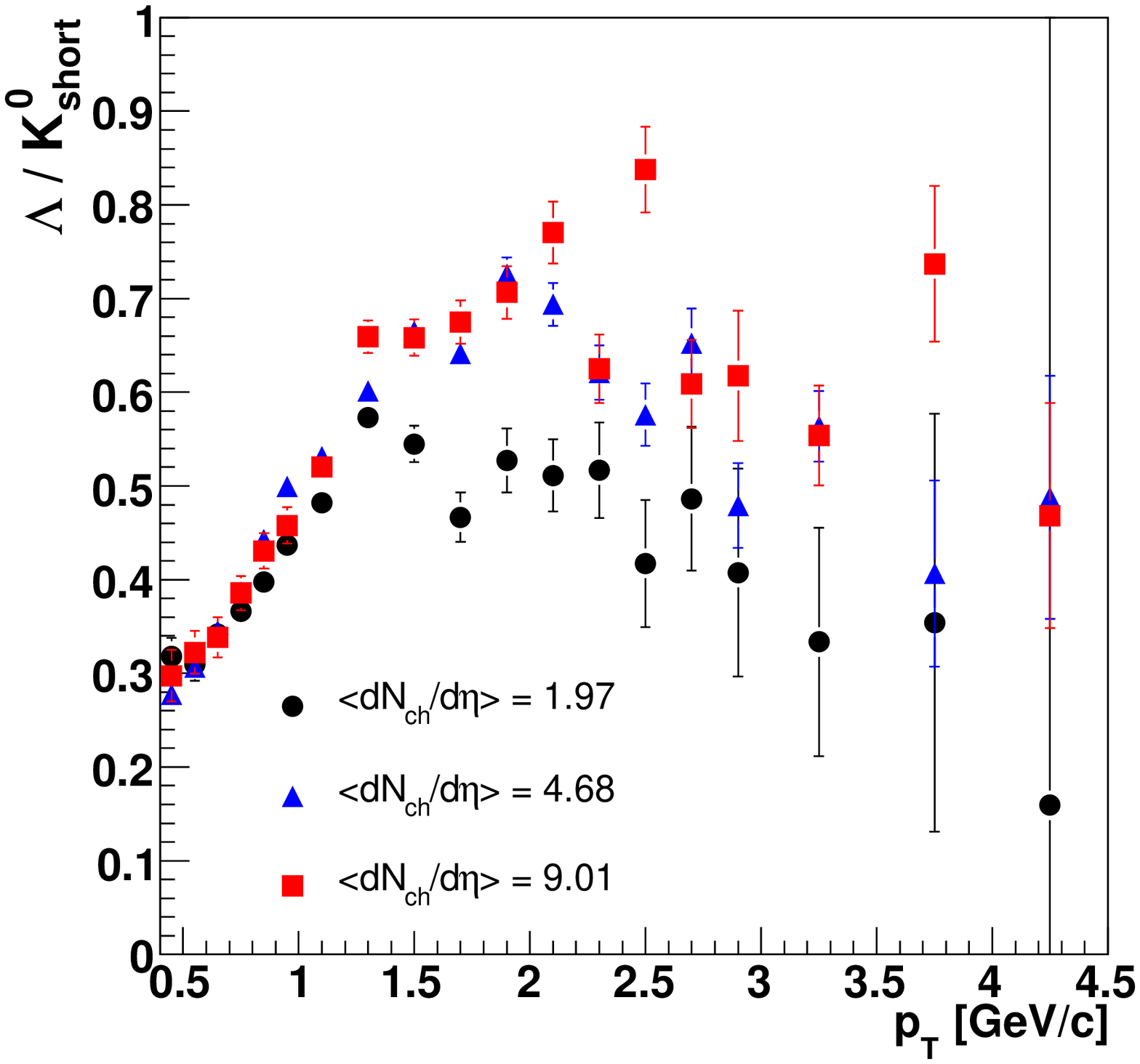}}
\caption{Ratios of multiplicity binned spectra to minimum bias spectra (R$_{pp}$) for 
\Ks~and $\Lambda$ and the ratio of the $\Lambda$ spectrum to the \Ks~spectrum in each 
multiplicity bin.  See text for further details.}
\label{fig:Rpp}
\end{figure*}
\begin{figure*}
\subfigure[\textsc{Pythia} for \Ks.]{
\label{fig:PythiaK0}
\includegraphics[width=0.67\columnwidth]{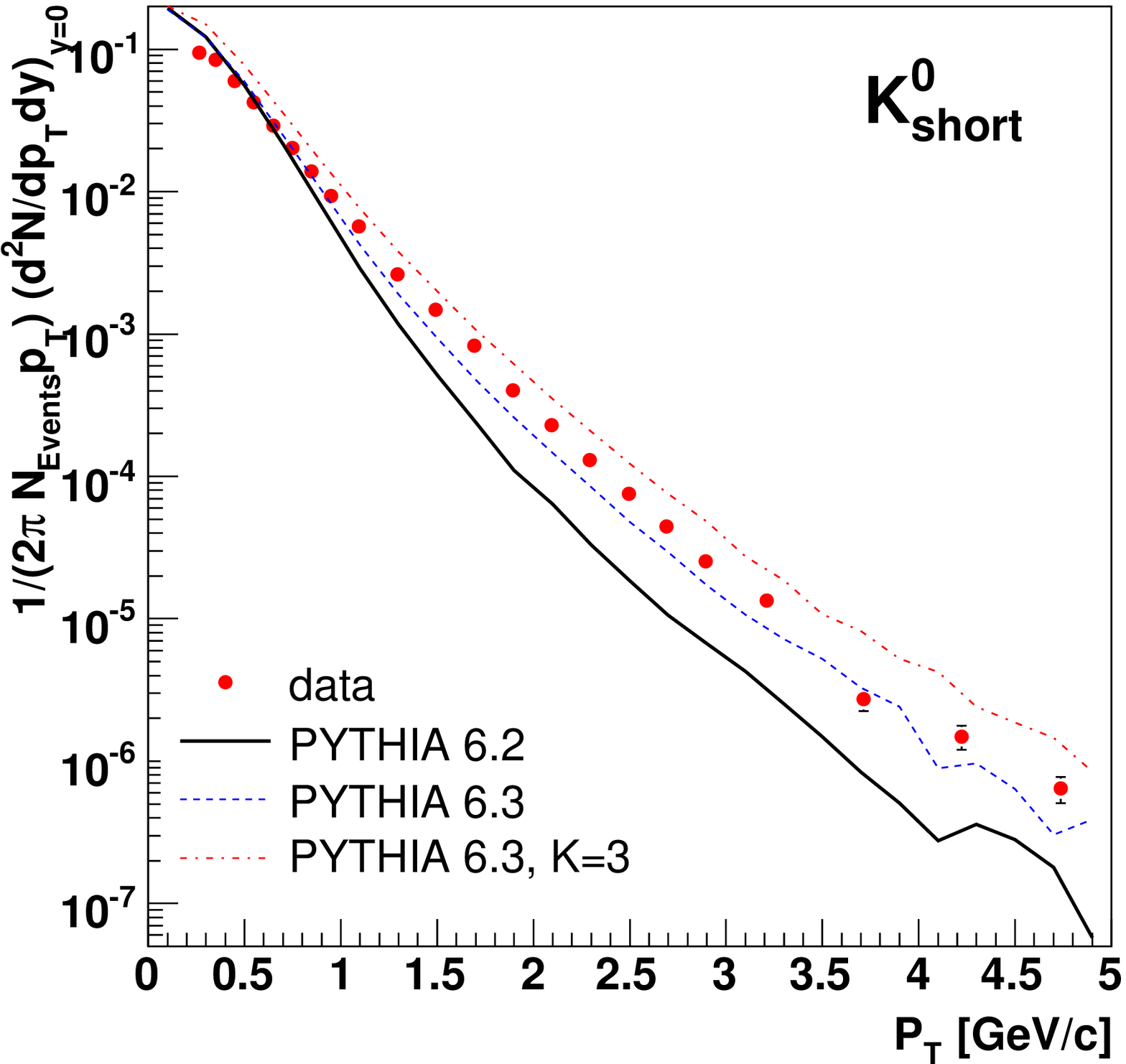}} 
\subfigure[\textsc{Pythia} for $\Lambda$.]{
\label{fig:PythiaLam}
\includegraphics[width=0.67\columnwidth]{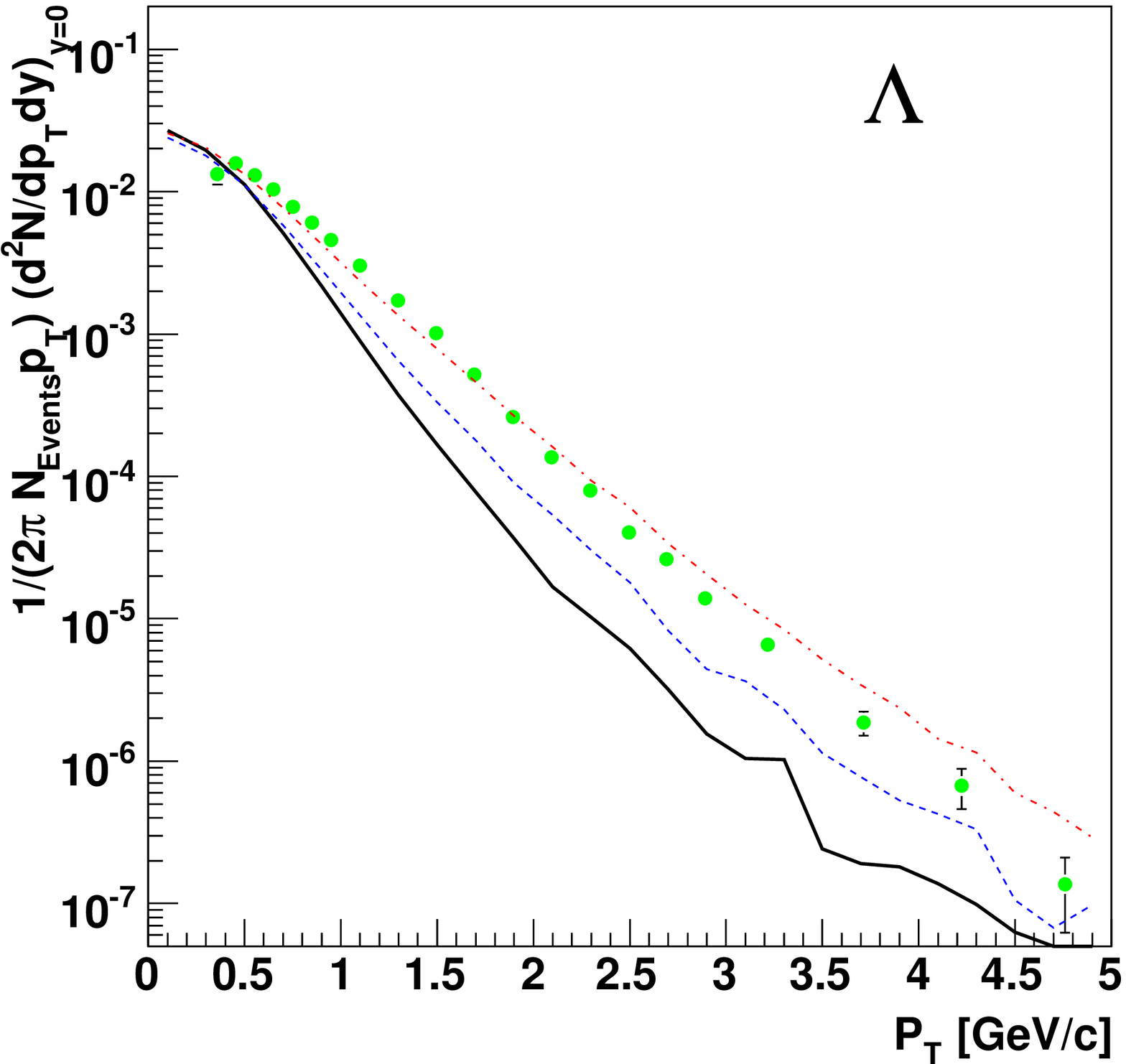}}
\subfigure[\textsc{Pythia} for $\Xi$.]{
\label{fig:PythiaXi}
\includegraphics[width=0.67\columnwidth]{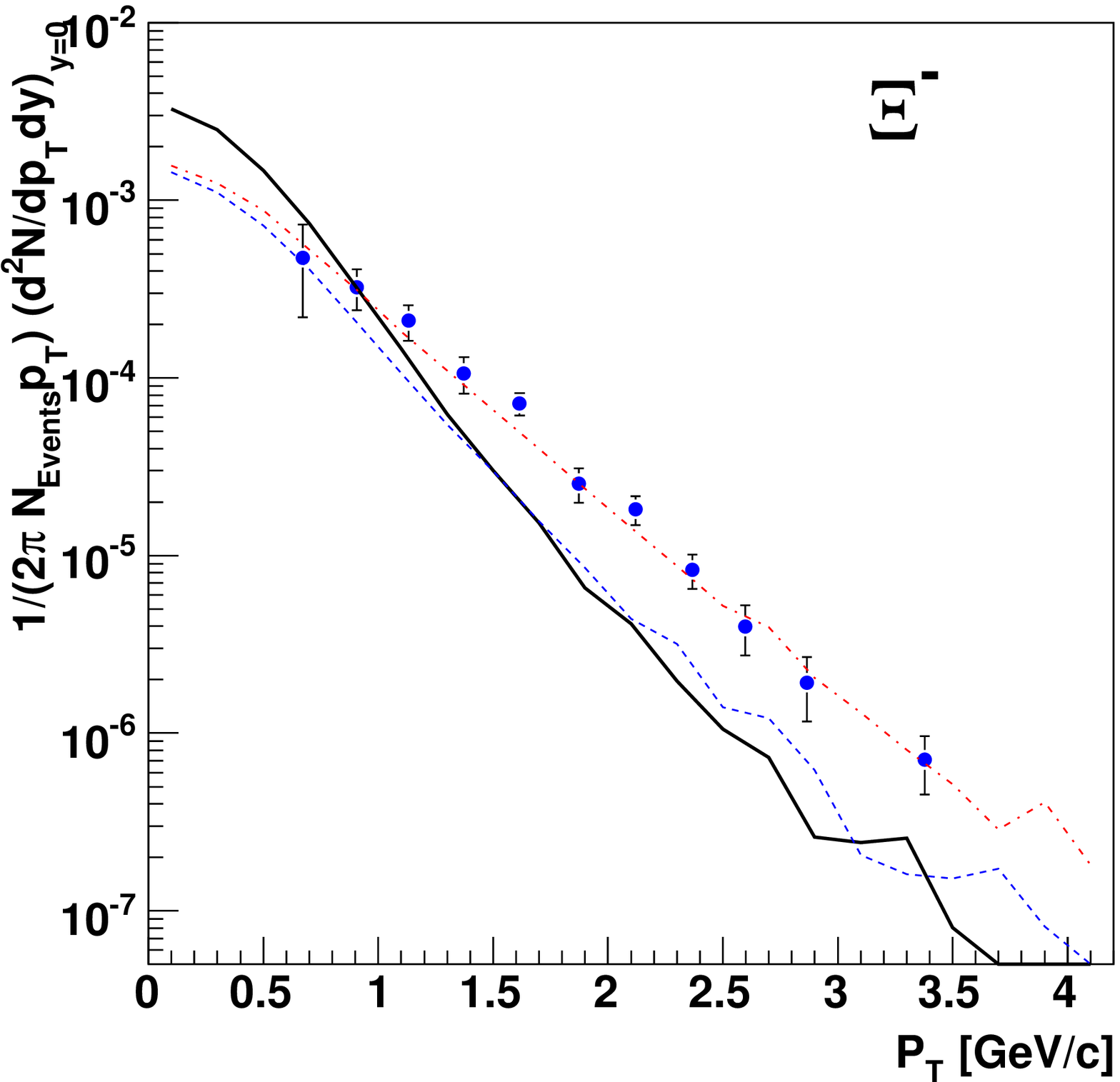}}
\caption{$\mathrm{K^{0}_{S}}$ (left), $\Lambda$ (center), and $\Xi^{-}$ (right) 
\pt~spectra compared to \textsc{Pythia}(ver 6.22 MSEL1, and ver 6.3) with the default 
K-factor=1 (solid and dashed curves respectively), and K-factor=3 (dot-dashed curve).}
\label{fig:pythiaSpec}
\end{figure*}

The scale difference is readily apparent but perhaps more interesting 
is the increasing trend of \mpt~with event multiplicity.  
The increase in \mpt~with multiplicity is faster for the $\Lambda$ than 
for the \Ks~and charged kaons over the range from 2 to 6 in $\langle dN_{ch}/d\eta \rangle$.  
The statistics available in the multiplicity-binned $\Xi+\overline{\Xi}$ do not allow 
a proper constraint of the L\'evy fit.  The points for $\Xi+\overline{\Xi}$ shown in Figure 
\ref{fig:mptmult} were determined from the error-weighted mean of the measured \pt~distribution 
only.  The present level of error on the $\Xi$ measurement does not allow a strong conclusion 
to be drawn, though the trend with increasing mass is suggestive.  This mass-ordering 
of the \mpt~multiplicity dependence has been observed in previous measurements at three different 
energies \cite{E735}.  In particular, the pions show little increase in \mpt~when going 
from low to high multiplicity collisions \cite{E735}.

Models inspired by pQCD such as \textsc{Pythia} suggest that the 
number of produced mini-jets (and thereby the event multiplicity) is correlated 
with the hardness ($Q^{2}$) of the collision.  
The effect of the mini-jets is to increase the multiplicity of the events
and their fragmentation into hadrons will also produce harder \pt~spectra.

The spectral shape cannot be characterized by a single number.  It is 
also possible to compare the multiplicity-binned spectra directly.  
We show in Figures \ref{fig:RppK0} and \ref{fig:RppLam} the ratio (R$_{pp}$) of the 
multiplicity-binned \pt~spectra to the multiplicity-integrated (minimum bias) spectra 
scaled by the mean multiplicity for each bin (see Equation \ref{equ:Rpp}) for \Ks~and 
$\Lambda$ respectively.  

\begin{equation}
R_{pp}(\text{p}_{\text{T}}) = F_{\text{scale}} \cdot \frac{dN/d\text{p}_{\text{T}}(\text{mult},\text{p}_{\text{T}})}
{dN/d\text{p}_{\text{T}}(\text{minbias},\text{p}_{\text{T}})},
\label{equ:Rpp}
\end{equation}
where
\begin{equation}
F_{\text{scale}} \equiv 
\frac{N_{events}(\text{minbias}) \cdot <N_{ch}(\text{minbias})>}{N_{events}(\text{mult}) \cdot <N_{ch}(\text{mult})>}.
\label{equ:Fscale}
\end{equation}

The changes in incremental shape from one multiplicity bin to 
the next then become easier to see.  The striking change in spectral shape going from 
the lowest to highest multiplicity bin is further evidence of the increasing contribution of 
hard processes (jets) to the high \pt~part of the spectra in high multiplicity events. 

Figure \ref{fig:LamK0Ratio} shows the $\Lambda$/\Ks~ratio as a function 
of \pt~in the various multiplicity bins.  We see in all three bins that the 
$\Lambda$ shows a sharper increase with \pt~in the low \pt~($\lesssim$ 1.5 GeV/$c$) 
part of the spectrum.  Furthermore there seems to be a relative increase in the 
$\Lambda$ production in the intermediate $1.5 \leq\mathrm{p_{T}}\leq 4.0$
 GeV/$c$ region.

\section{Model Comparisons \label{models}}

\begin{figure*}[ht]
\epsfig{figure=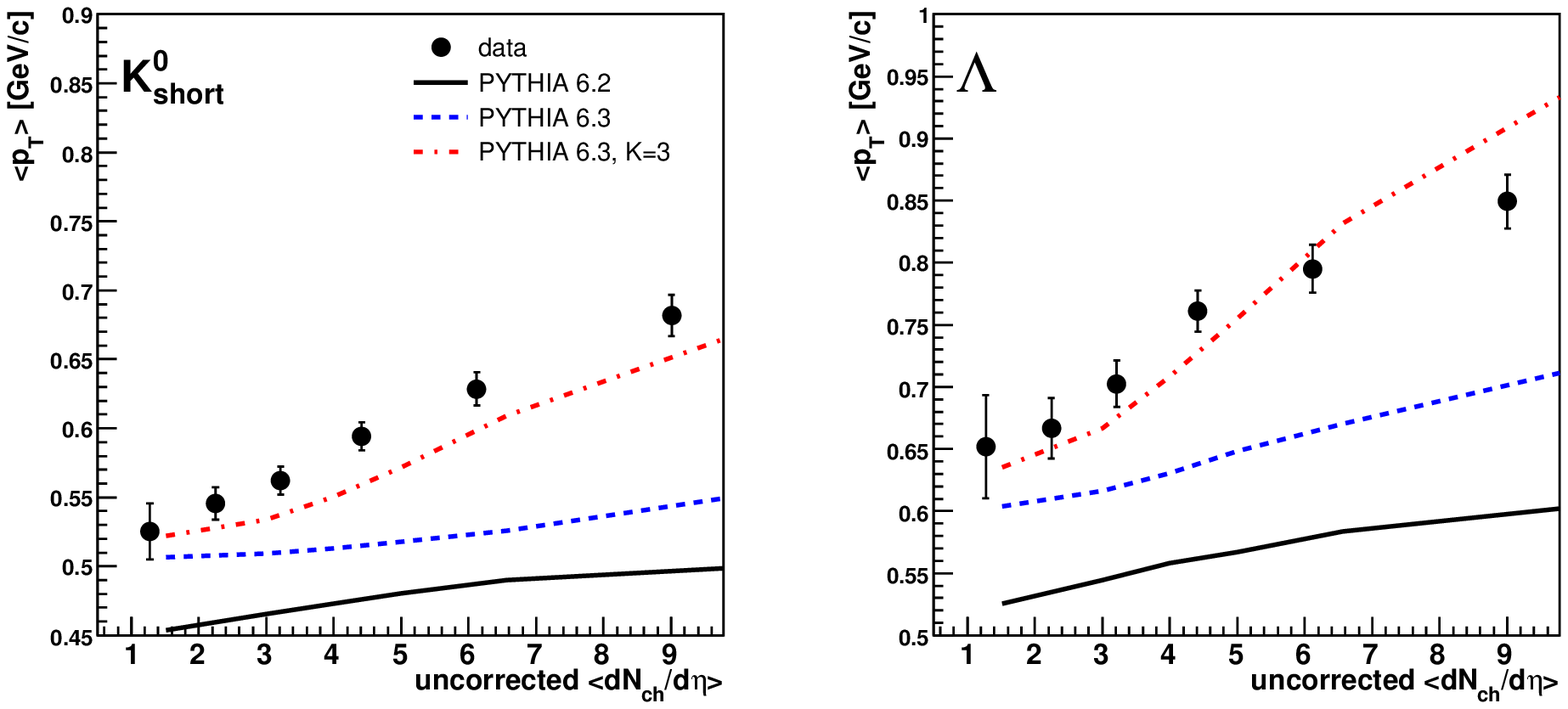,width=0.7\linewidth}
\caption{$\mathrm{K^{0}_{S}}$ (left) and $\Lambda$ (right) multiplicity-binned
\mpt~compared to \textsc{Pythia}(ver 6.22 MSEL1, and ver 6.3) with default 
K-factors (solid and dashed curves respectively), and K-factor=3 (dot-dashed curve).}
\label{fig:pythiaMpt}
\end{figure*}

\subsection{Comparison to PYTHIA (LO pQCD)}

At the present time, the most ubiquitous model available for the
description of hadron+hadron collisions is the \textsc{Pythia} 
event generator.  \textsc{Pythia} was based on the Lund
string fragmentation model \cite{string1,string2} but has been
refined to include initial and final-state parton showers and many
more hard processes.  \textsc{Pythia} has been shown to be successful
in the description of collisions of $\mathrm{e}^{+}\mathrm{e}^{-}$,
$p+\overline{p}$ and fixed target $p+p$ systems (see for example,
ref. \cite{Field}).

In this paper we have used \textsc{Pythia} v6.220 and v6.317
(using default settings with in-elastic cross-section (MSEL=1)) in order
to simulate \pt~spectra for \Ks, $\Lambda$ and $\Xi$.  These have then been 
compared with the measured data.

As shown in Figure \ref{fig:pythiaSpec}, although there is some
agreement at low \pt, there are notable differences above \pt $\sim
$1.0 GeV/$c$, where hard processes begin to dominate.
\textsc{Pythia} v6.2 underestimates the $\Lambda$ yield by almost an
order of magnitude at \pt=3 GeV/$c$.  With the newer version 6.3,
released in January 2005, these large discrepancies have been largely 
reconciled for \Ks~but remain significant for $\Lambda$ and $\Xi$.
This version includes a significantly modified description of the 
multiple parton scattering processes.  The red dot-dashed lines in Figure 
\ref{fig:pythiaSpec} represent a simple tune that was done with \textsc{Pythia} 
6.317 which will be described in more detail below.

To try and understand the difference between \textsc{Pythia}
and our results, we made comparisons of \mpt~versus uncorrected charged
multiplicity for \Ks~and $\Lambda$, as shown in Figure \ref{fig:pythiaMpt}.
As expected from the previous figure, version 6.2 fails to reproduce
the minimum bias magnitude of \mpt.  Although version 6.3 is capable of
reproducing our minimum bias values of \mpt~it clearly fails to reflect
its increase with charged multiplicity, suggesting that further
tuning is necessary.
\begin{figure*}
\subfigure[~Quark jets.]{
\label{fig:PythiaQuark}
\includegraphics[width=0.67\columnwidth]{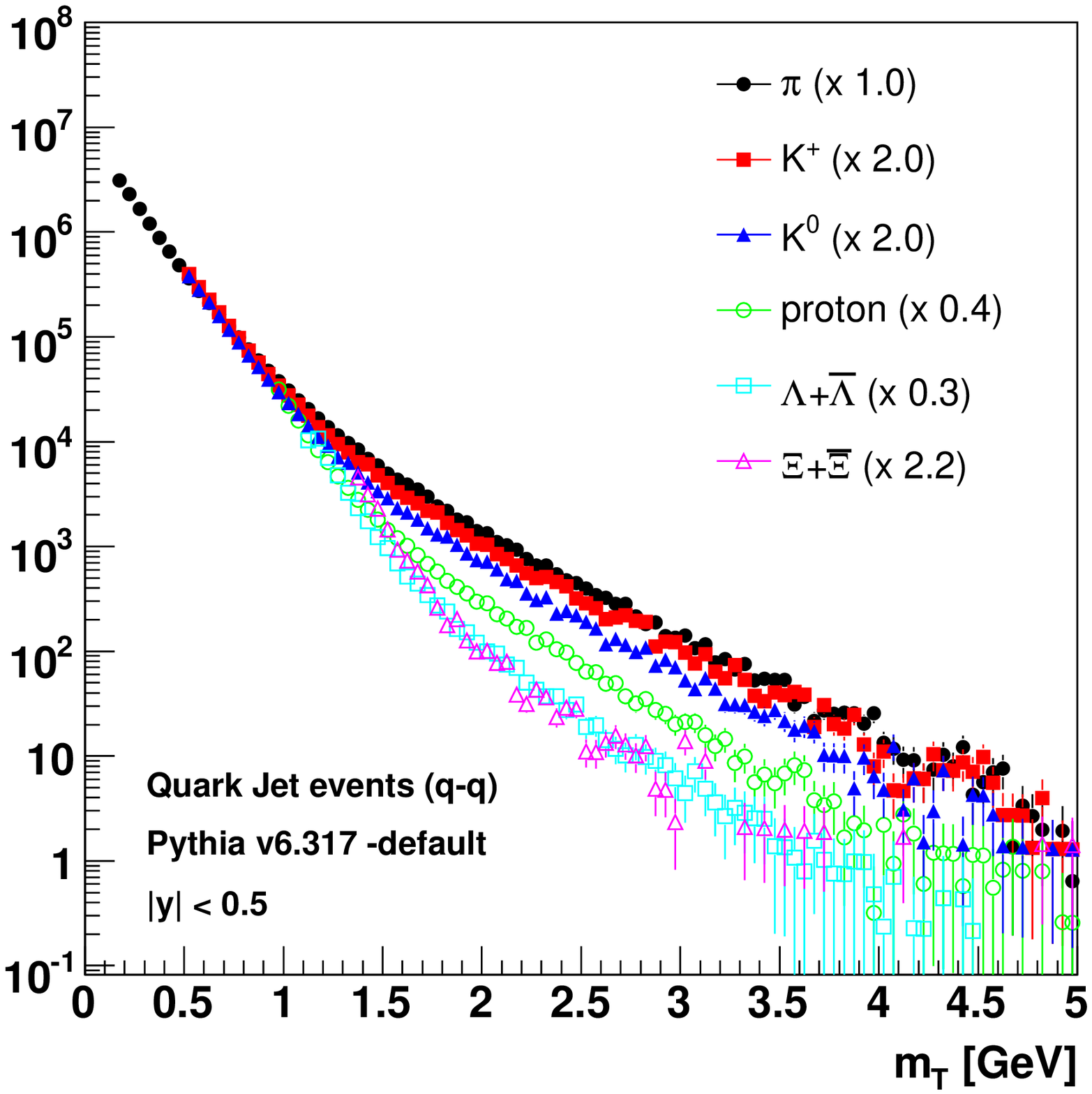}} 
\subfigure[~Gluon jets.]{
\label{fig:PythiaGluon}
\includegraphics[width=0.67\columnwidth]{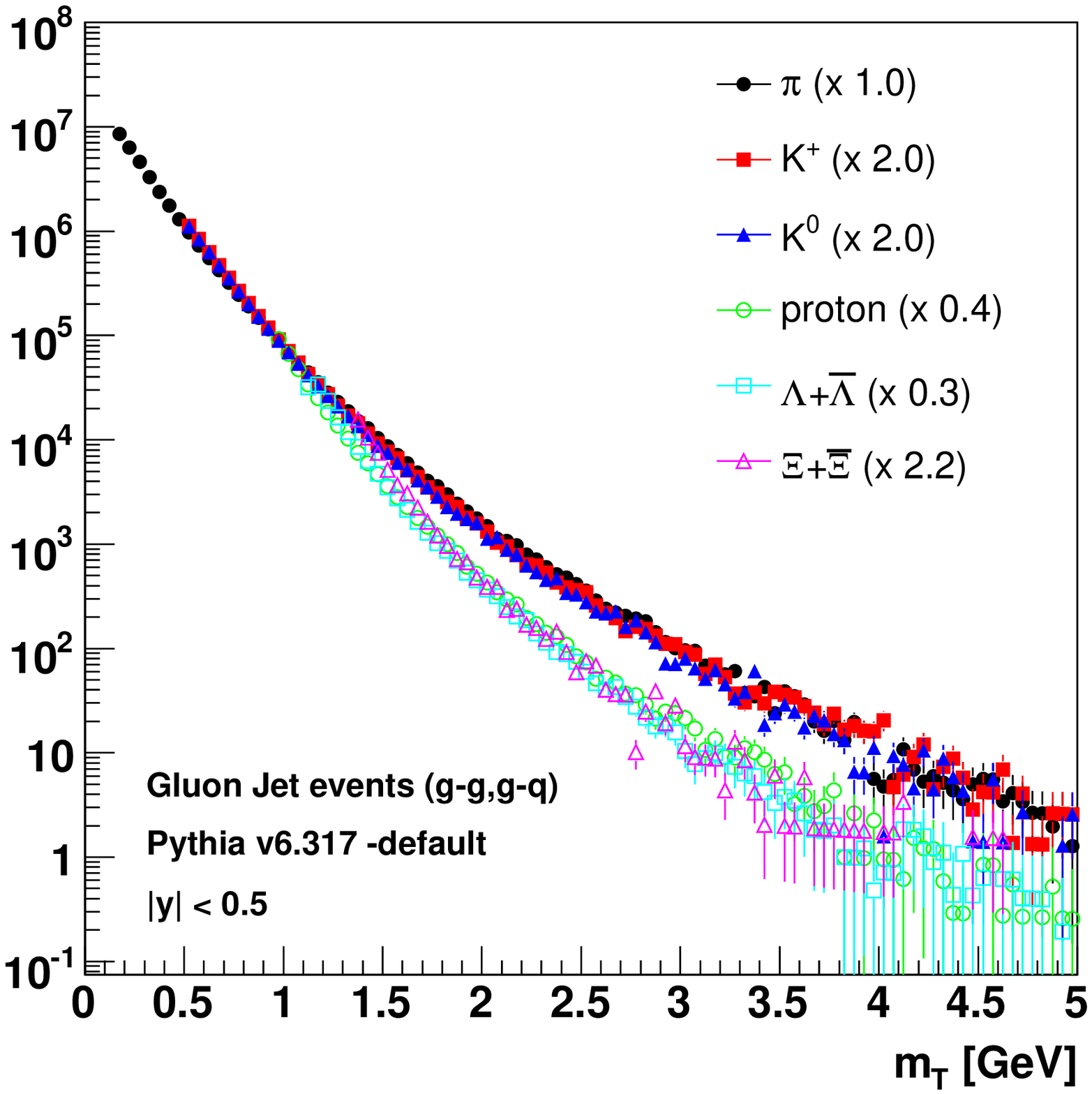}}
\subfigure[~All.]{
\label{fig:PythiaAll}
\includegraphics[width=0.67\columnwidth]{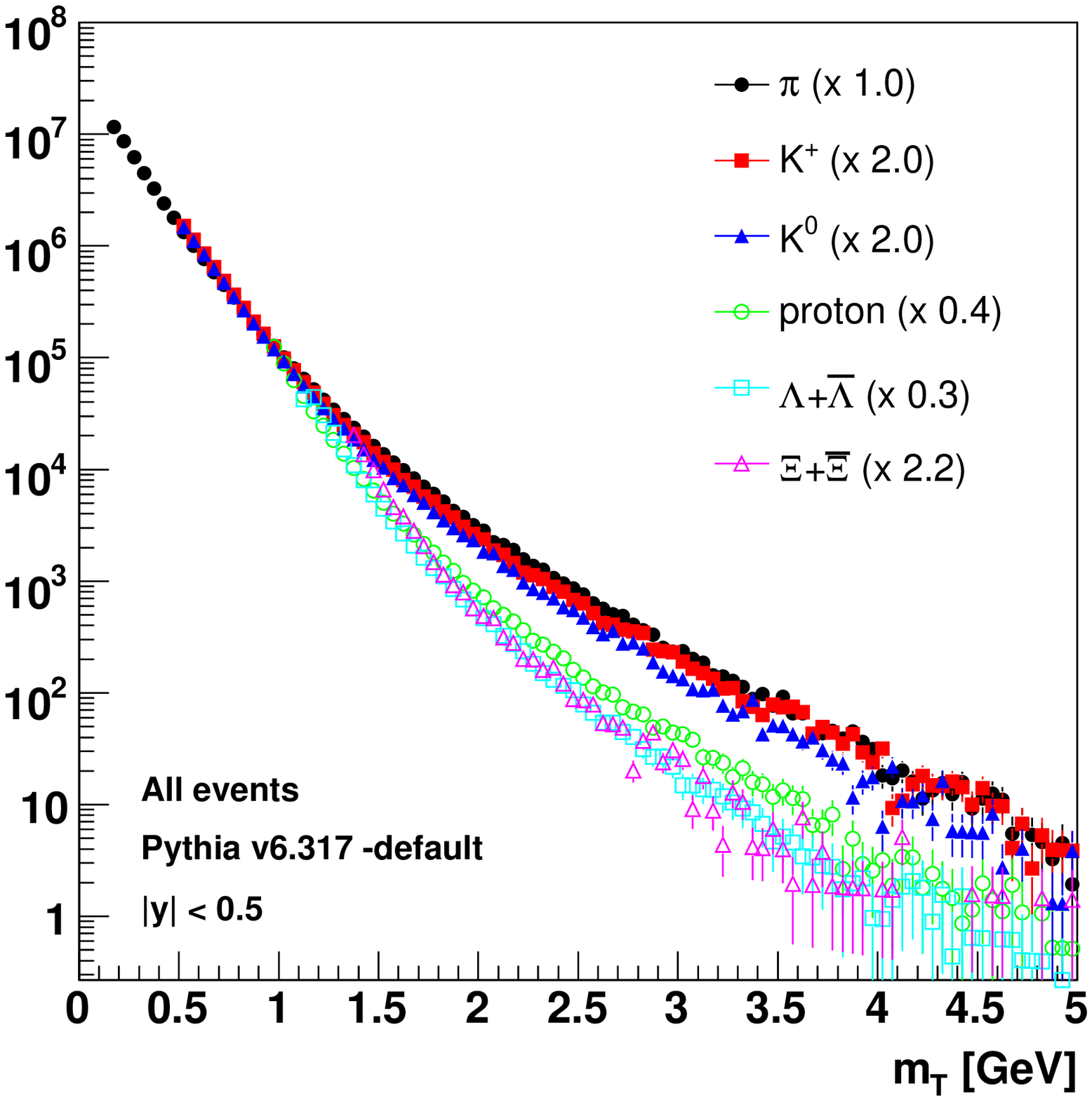}}
\caption{\mt-scaling results from \textsc{Pythia} 6.3 with default settings.  The quark or gluon jet 
selections are based on the final state partons being qq or $gg$ respectively.  The second panel (\ref{fig:PythiaGluon}) also contains mixed final states (q$g$).  The spectra have been scaled by the factors listed in the legends.  
See text for more details.}
\label{fig:PythiaMtScaling}
\end{figure*}

In order to improve the agreement with our data we have made some
simple changes to the \textsc{Pythia} default parameters.  In
particular, by increasing the K-factor to a value of 3 (set to 1 in
the defaults) there is an enhancement of the particle yield at high
\pt~in the model which allows it to better describe the data.

The K-factor, which represents a simple factorization of
next-to-leading order processes (NLO) in the \textsc{Pythia} leading
order (LO) calculation, is expected to be between 1.5--2 for most
processes, such as Drell-Yan and heavy quark production
\cite{kfactor} at higher energies.  Based on these measurements, a
K-factor of 3 would signal a large NLO contribution,
particularly for light quark production at RHIC energies.
Intriguingly, a large K-factor has been estimated for the $\sqrt{s}
\sim$ 200 GeV regime at RHIC based on the
energy dependence of charged hadron spectra \cite{eskola-kfactor}.
So it seems that for light quark production at lower energies, 
NLO contributions are important and a comparison of our data to 
detailed pQCD based NLO calculations is more appropriate.

With the addition of this K-factor, we can see that the \pt~spectra
for $\Lambda$ and $\Xi$ in Figure \ref{fig:pythiaSpec} agree even
better with the model, with the \Ks~data falling slightly below the prediction.  
More importantly, the \textsc{Pythia} results of \mpt~versus charged
multiplicity, including the enhanced K-factor, are now in much better
agreement with the data, as seen in Figure \ref{fig:pythiaMpt}.

Figure \ref{fig:PythiaMtScaling} shows the results of separating 
\textsc{Pythia} events based on their final state parton content.  
Events where the final state is qq are labeled as containing 
``quark jets" while events with $gg$ are labled as containing 
gluon jets.  Figure \ref{fig:PythiaQuark} shows that events with only 
quark-jet final states seem to show a mass splitting in the high 
\mt~region while events whose final states contain jets from gluons (Figure 
\ref{fig:PythiaGluon}) show a shape difference between mesons and baryons with 
the meson spectra being harder than the baryon spectra.  The shape difference is 
also apparent in Figure \ref{fig:PythiaAll} which contains all final states 
including those with both quark and gluon jets.  This shape difference  
could be simply related to the fact that a fragmentation process could 
impart more momentum to a produced meson than a produced baryon based 
on mass and energy arguments.  This taken together with the results shown 
in Sections \ref{mt-scalingSection} and \ref{Ratios} indicates that above 2 GeV in 
\textit{transverse mass}, the spectra contain significant contributions 
from \textit{gluon}-jet fragmentation rather than quark-jet fragmentation.
\begin{figure}[ht]
\epsfig{figure=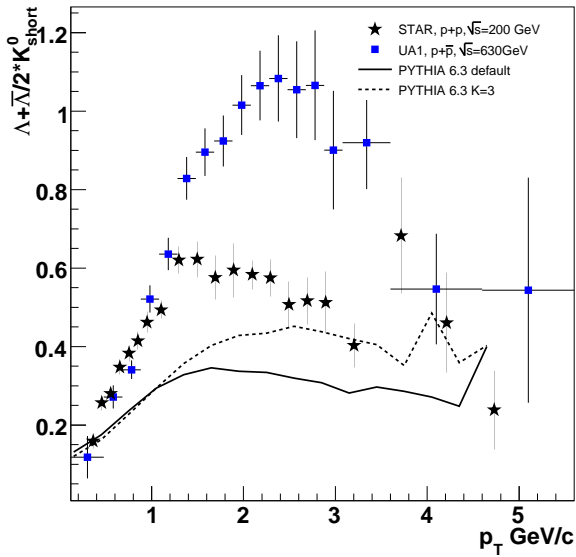,width=\columnwidth}
\caption{$\Lambda$/\Ks~as a function \pt~compared with \textsc{Pythia}.}
\label{fig:LamK0Comp}
\end{figure}

In Figure \ref{fig:LamK0Ratio} we showed the $\Lambda$ to \Ks~ratio 
separated into multiplicity bins.  Figure \ref{fig:LamK0Comp} shows 
the multiplicity integrated ratio compared with \textsc{Pythia} 
calculations using the default settings as well as a K-factor of 3.  
Here we see again the same shape difference between the $\Lambda$ and 
the \Ks~that is seen for baryons and mesons in general in Figure 
\ref{fig:mtscaling} and in the p/$\pi$ ratio \cite{STAR-rDEDX}.  
\textsc{Pythia} is not able to reproduce the full magnitude of the 
effect in either ratio \cite{STAR-rDEDX}.  The $\Lambda$ to \Ks~ratio shows a 
similar shape in $\sqrt{\text{s}}=200$ GeV Au+Au collisions though the 
magnitude is larger and multiplicity-dependent \cite{Lamont06}.  Also, measurements 
from UA1 at $\sqrt{\text{s}}=630$ GeV indicate the magnitude may also 
be dependent on beam energy \cite{Bocquet96}.
\begin{figure*}[ht]
\epsfig{figure=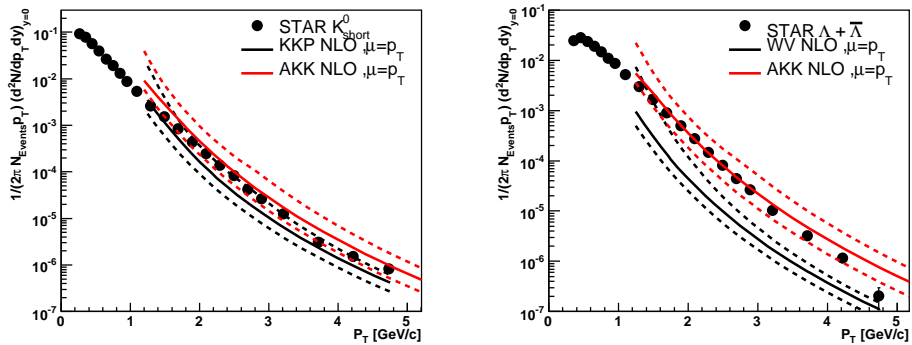,width=1.5\columnwidth}
\caption{$\mathrm{K^{0}_{S}}$ (left) and $\Lambda$ (right) particle spectra
(circles) compared to Next-To-Leading-Order (NLO calculations) by Werner
Vogelsang based on specific \Ks~\cite{KKP} and $\Lambda$ \cite{Vogelsang}
 fragmentation functions.  Dashed lines illustrate the uncertainty 
due to the choice of factorization scale.}
\label{fig:NLOComp}
\end{figure*}

\subsection{Comparison to NLO pQCD Calculations}

In Figure \ref{fig:NLOComp} we compare the \Ks~and $\Lambda$ spectra to 
NLO pQCD calculations including fragmentation functions for the \Ks~from 
Kniehl, Kramer, and P\"otter (KKP) \cite{KKP} and a calculation by 
DeFlorian, Stratmann, and Vogelsang for the $\Lambda$ \cite{Vogelsang}.  The variations 
in $\mu$ show the theoretical uncertainty due to changes of the factorization 
and renormalization scale used.  The factorization and renormalization
scale allows one to weight the specific hard scattering contributions of the parton densities 
to the momentum spectrum.  Although for the \Ks~reasonable agreement is achieved 
between our data and the pQCD calculation, the comparison is much less favorable for the $\Lambda$.  
Considering that good agreement was achieved for charged pion \cite{STAR-rDEDX} 
and $\pi^{0}$ \cite{phenix-pi0,STAR-pi0} spectra and yields at the same energy, 
our comparison and the comparisons in \cite{STAR-rDEDX} suggest that the region of agreement 
with NLO pQCD calculations may be particle species dependent.  The baryons are 
more sensitive to the gluon and non-valence quark fragmentation function, which 
is less constrained at high values of the fractional momentum $z$ \cite{Bourrely Soffer}.

Recently, the OPAL collaboration released new light quark flavor-tagged 
$e^{+}e^{-}$ data which allows further constraint of the fragmentation functions 
\cite{OPAL}.  Albino-Kniehl-Kramer (AKK) show that these flavor separated fragmentation
functions can describe our experimental data better \cite{AKK}.  However, in order to
achieve this agreement, AKK fix the initial gluon to $\Lambda$ fragmentation function
$(D_{g}^{\Lambda})$ to that of the proton $(D_{g}^{p})$, and apply an additional 
scaling factor.  They then check that this modified $D_{g}^{\Lambda}$ 
also works well in describing the $\overline{p}+p$ \sps~data at \sqs=630 GeV.  
So, it appears that the STAR data is a better constraint for the high $z$ part 
of the gluon fragmentation function than the OPAL data.  Similar conclusions have 
been drawn elsewhere with respect to the important role of RHIC energy 
$p+p$ collisions \cite{Fai}.  Recent studies of forward $\pi^{0}$ production also 
suggest that the region of agreement with NLO calculations extends as far out 
as 3.3 units in $\langle\eta\rangle$ \cite{FPD}.

\subsection{Comparison to EPOS}

Finally we compare our data to the version 1.02 of the EPOS model
\cite{EPOSDetails}.  This model generates the majority of
intermediate momentum particles by multiple parton interactions in
the final state rather than fragmentation.  The multi-parton cross section 
is enhanced through a space-like parton cascade in the incoming parton systems.  
The outgoing, time-like parton emission, is allowed to self-interact and to 
interact with the di-quark remnants.  The interactions can be either 
elastic or inelastic.  The overall result is a strong probability for 
multi-parton interactions before hadronization.  The cascades are modeled through 
so-called parton ladders which also include multiple scattering contributions of the
di-quark remnants from a hard parton scattering in a $p+p$ collision.
Furthermore, by taking into account the soft pomeron interactions, the
model is able to describe the $p+p$ spectra down to low \pt.
Finally, the inclusion of parton ladder splitting in asymmetric d+Au
collisions yields a good description of the difference between $p+p$
and d+Au spectra in the same theoretical framework.  Further details
of the model can be found elsewhere \cite{EPOSDetails}.
\begin{figure*}[ht]
\epsfig{figure=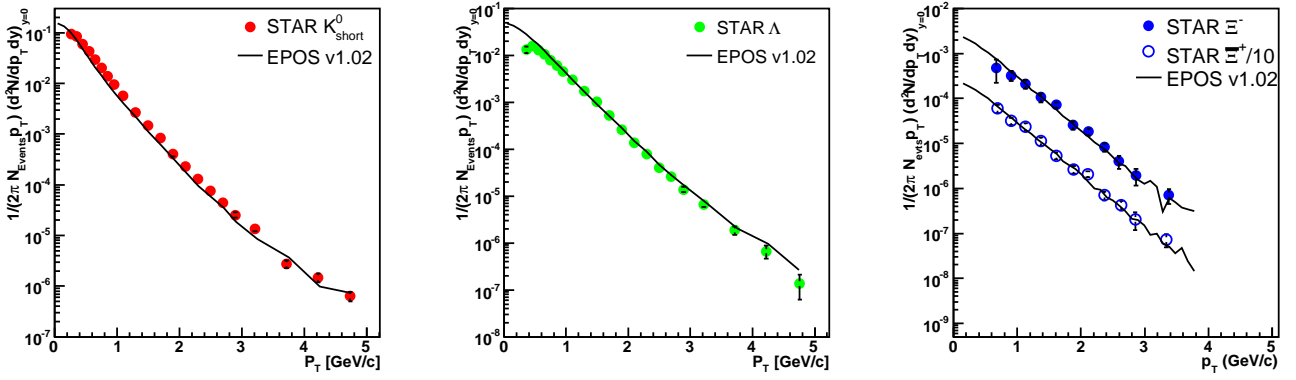,width=\linewidth}
\caption{Comparison of \Ks~, $\Lambda$, $\Xi^{-}$, and $\overline{\Xi}^{+}$
spectra with calculations from EPOS v1.02.}
\label{fig:eposComp}
\end{figure*}

EPOS shows remarkable agreement with BRAHMS, PHENIX and STAR data
for pion and kaon momentum spectra and \mpt~in $p+p$ and d+Au collisions at both 
central and forward rapidities (\cite{STAR-rDEDX,EPOS-RHIC,EPOSDetails} and references therein).  
Figure \ref{fig:eposComp} shows that this trend also continues for the heavier strange
particles at mid-rapidity.  The agreement in $p+p$ collisions in the measured 
\pt~region is largely due to a strong soft component from string
fragmentation in the parton ladder formalism.  Remnant and hard
fragmentation contributions are almost negligible at these moderate
momenta.  The soft contribution dominates the kaon spectrum out to 1 GeV/$c$
and the $\Xi$ spectrum out to 3 GeV/$c$.  As the momentum
differences between (di-quark,anti-di-quark) and (quark,anti-quark) string
splitting are taken into account, and the current mass difference
between light and strange quarks is folded into the spectral shape,
a comparison between the spectra exhibits a flow-like mass
dependence.

The agreement with EPOS is better than even the best NLO calculations.
A detailed discussion of the differences between EPOS and NLO
calculations is beyond the scope of this paper, but it should be
mentioned that the two models are, in certain aspects, complementary.
More measurements of a) heavier particles and b) to much
higher \pt~are needed in order to distinguish between the different
production mechanisms.  In summary, the data show the need for
sizeable next-to-leading-order contributions or soft multi-parton
interactions in order to describe strange particle production in
$p+p$ collisions.

\subsection{Statistical Model}

The application of statistical methods to high energy hadron-hadron
collisions has a long history dating back to Hagedorn in the 1960s
\cite{Hagedorn65,Hagedorn68a,Hagedorn68b}.  Since then statistical
models have enjoyed much success in fitting data from relativistic
heavy-ion collisions across a wide range of collision energies 
\cite{Hagedorn70,Siemens79,Mekjian82,Csernai86,Stocker86,Cleymans93,Rafelski02,PBM03}.  
The resulting parameters are interpreted in a thermodynamic sense, allowing
a ``true'' temperature and several chemical potentials to be ascribed to
the system. More recently, statistical descriptions have been applied to
$p+p$ and  $\overline{p}+p$ collisions \cite{Becattini}, and
even $e^{+}+e^{-}$ \cite{Becattinie+e-}, but it remains unclear as to
how such models can successfully describe particle production and kinematics
in systems of small volume and energy density compared to heavy-ion
collisions.

It is important to note that a $p+p$ system does not have to be \textit{thermal}
on a macroscopic scale to follow statistical emission.  For example, 
Bourrely and Soffer have recently shown that jet fragmentation can be parametrized
with statistical distributions for the fragmentation functions and parton
distribution functions \cite{Bourrely Soffer}.  In this picture, the
apparently statistical nature of particle production observed in our data
would be a simple reflection of the underlying statistical features of
fragmentation.  It is interesting to note that Biro and Mueller have shown
that the folding of partonic power law spectra can produce exponential 
spectral shapes of observed hadrons in the intermediate \pt~region with 
no assumption of temperature or thermal equilibrium whatsoever \cite{Biro}.  

Another possibly related idea is that of \textit{phase space dominance} in 
which all possible final state configurations (\textit{i.e.} those that are consistent with 
the energy, momentum, and quantum numbers of the initial state) are populated 
with equal probability \cite{Hormuzdiar}.  The finite energy available in the 
collision allows many more final state configurations that contain low mass particles 
than high mass particles.  The final state configurations containing high 
mass particles are therefore less likely to be observed not because they 
are less probable, but because there are fewer of them relative to the 
low mass configurations.
\begin{figure}[ht]
\epsfig{figure=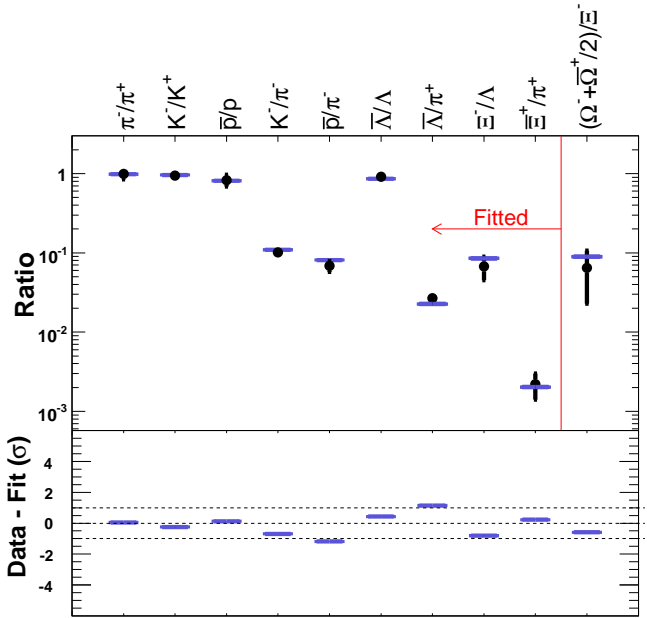,width=\columnwidth}
\caption{Parameters of ratio data to statistical model fit using THERMUS.  Filled 
circles are ratios from $\sqrt{\text{s}}$=200 GeV collisions in STAR.  Solid 
lines are the results from the statistical model fit.  All ratios to the left of the 
vertical line were used in the fit.  The $(\Omega^{-}+\overline{\Omega}^{+}/2)/\Xi^{-}$ ratio was 
then predicted from the fit results.  The dashed lines in the 
lower panel are guides for the eye at 1 $\sigma$.}
\label{fig:thermus}
\end{figure}

We include in this section the results of a canonical statistical model fit, using
THERMUS \cite{THERMUS}, to the STAR feed-down corrected ratios from
$p+p$ collisions at $\sqrt{s}$=200 GeV.  We used only the canonical formalism 
as it has been determined from a micro-canonical calculation that the volume of 
$p+p$ collision systems does not exceed 100 fm$^3$ \cite{Liu04}.  Previous results 
have shown that such a small volume invalidates the use of a grand-canonical 
treatment \cite{Liu03}.  The canonical calculation involves only the temperature (T), 
baryon number (B), charge (Q), strangeness saturation factor ($\gamma_{S}$), and 
the radius.  For this fit, B and Q were both held fixed at 2.0.  The resulting 
parameters are presented in Table \ref{tab:thermus} 
and a graphical comparison is presented in Figure \ref{fig:thermus}.

\begin{table}[h]
\begin{center}
\setlength\extrarowheight{2pt}
\begin{tabular}{|c|c|}
\hline  &
\textbf{Canonical Value} \tabularnewline
\hline \textbf{T}  &
        0.1680 $\pm$ 0.0081 GeV \tabularnewline  
\hline $\mathbf{B}$  &
        2.000 (fixed) \tabularnewline  
\hline $\mathbf{Q}$  &
        2.000 $e$ (fixed) \tabularnewline  
\hline $\mathbf{\gamma_{S}}$  &
        0.548 $\pm$ 0.052 \tabularnewline  
\hline $\mathbf{radius}$  &
        3.83 $\pm$ 1.15 fm \tabularnewline  
\hline
\end{tabular}
\caption{Comparison of a canonical fit to the STAR
feed-down corrected ratios from $p+p$ collisions at $\sqrt{s}$=200 GeV.  
The $\chi^{2}/ndf$ of the fit was 4.14 / 6 = 0.69.  See text for further details.}
\label{tab:thermus}
\end{center}
\end{table}

The interpretation of the fit parameters is difficult in the context of 
a $p+p$ collision where the system is not expected to thermalize and the 
volume is small.  It is important to note that in a pure thermal model, 
all emitted particles would be expected to reflect the same temperature.  
Non-thermal effects such as flow would modify this result.  In $p+p$ collisions, 
the particle spectra clearly show different slopes and those slopes are 
not in agreement with the T parameter that results from the statistical 
model fit to the particle ratios.  As no flow is thought to be present in 
the $p+p$ system and the results of Section \ref{mt-scalingSection} support that 
conclusion, this result is a further indication of contributions to 
the particle spectra from non-thermal processes like mini-jets.

\section{Summary and Conclusions \label{Summary}}

We have presented measurements of \kp, \km, \Ks, $\Lambda$, $\overline{\Lambda}$, 
$\Xi^{-}$, $\overline{\Xi}^{+}$, and $\Omega^{-}+\overline{\Omega}^{+}$ \pt~spectra 
and mid-rapidity yields from $\sqrt{s}$=200 GeV $p+p$ collisions in STAR.  
Corrections have been made for detector acceptance and efficiency as well as the multiplicity 
dependence of the primary vertex finding and, in the case of the $\Lambda$ 
and $\overline{\Lambda}$, feed-down from higher mass weak decays.  It was found 
that the measured range of transverse momentum necessitates a  
functional form that accounts for the power-law like shape at high \pt.  
We have used a L\'evy function to fit the spectra and extrapolate to low \pt.  

The \mpt~and mid-rapidity yields are in excellent agreement for all species 
with previous measurements at the same energy but with greatly improved 
precision.  The anti-particle to particle ratios are flat with \pt~over 
the measured range for both the $\Lambda$ and $\Xi$ and therefore show 
no sign of quark-jet dominance at high \pt.  The \pt~integrated 
ratios approach unity with increasing strangeness content.  The anti-baryon 
to baryon ratios suggest that the mid-rapidity region at RHIC is almost baryon-free, 
at least in $p+p$ collisions.  The amount of deviation from unity expected from 
differing parton distribution functions must first be determined before any claim 
of significant baryon number transport from beam rapidity to mid-rapidity can be made.  

We have demonstrated the scaling of transverse mass spectra for low 
\pt~mesons and baryons onto a single curve to within 30\% out to 
approximately 1.5 GeV in \mt.  Above 2 GeV the \mt~spectra show a clear difference 
in shape between mesons and baryons with the mesons being harder than the baryons.  
This is the first observation of a difference between baryon and meson spectra 
in $p+p$ collisions and is mainly due to the high \pt~(and therefore high \mt) 
coverage of the strange particles presented here.  \textsc{Pythia} 6.3 seems to 
account for this effect and suggests it is mostly due to the dominance of gluon 
jets.  More data are needed to determine the range of the effect.

The mean transverse momentum as a function of particle mass 
from both the $p+p$ and Au+Au systems has been compared.  Both systems show 
a strong dependence of \mpt~on particle mass.  It is also worth noting that 
the mass-dependence of \mpt~in the $p+p$ system seems to be independent 
of collision energy as the parameterization of the $\sqrt{s}=25$ GeV ISR data 
seems to work well over the same range of measured masses at RHIC.

The dependence of \mpt~on event multiplicity was also 
studied for each of the three species (and anti-particles).  The \mpt~shows 
a clear increase with event multiplicity for the \Ks~and $\Lambda$ particles.  
There may be a mass-ordering to the increase as the 
$\Lambda$ baryons show a slightly faster increase with multiplicity than the 
\Ks, but the present level of error on the $\Xi$ data does not allow a 
definite statement to be made. 

The multiplicity-binned \Ks~and $\Lambda$ spectra show a clear 
correlation between high multiplicity events and the high \pt~parts 
of the spectra.  The spectral shapes for the \Ks~and $\Lambda$ are observed 
to change with event multiplicity and the $\Lambda$ to \Ks~ratio 
increases over the lower \pt~range and reaches higher values in the 
\pt~range above $\sim$1.5 GeV/$c$ for larger multiplicites.  This suggests that 
the high multiplicity events produce more $\Lambda$ hyperons relative to \Ks~than the 
low multiplicity events.

Comparisons of our spectra with \textsc{Pythia} v6.221 show only poor 
agreement at best without adjustment of the default parameters.  In 
the relatively high \pt~region (above 2 GeV/$c$) there is nearly an order 
of magnitude difference between our data and the model calculation.  The 
more recent \textsc{Pythia} 6.3 provides a much better description of our 
\Ks~data though a K-factor of 3 is required to match the $\Lambda$ and 
$\Xi$ spectra as well as the observed rate of increase of \mpt~with 
multiplicity.  NLO pQCD calculations with varied factorization scales are able 
to reproduce the high \pt~shape of our \Ks~spectrum but not the $\Lambda$ 
spectrum.  Previous calculations at the same energy have been able to match 
the $\pi^{0}$ spectra almost perfectly, which suggests that there may be a 
mass dependence to the level of agreement achievable with pQCD.

The EPOS model has previously provided excellent descriptions of the 
\pim, \km, and proton spectra from both $p+p$ and d+Au collisions 
measured by BRAHMS, PHENIX, and STAR at mid-rapidity and forward rapidity.  We 
have extended the comparison to strange and multi-strange mesons and baryons 
and have found the agreement between our data and the EPOS model to be at 
least as good as the best NLO calculations.

We have demostrated the ability of the statistical model to fit our data 
to a reasonable degree with three parameters.  Interpretation of the resulting 
parameters in the traditional fashion is not possible as the $p+p$ colliding system is 
not considered to be thermalized.  The T parameter does not agree with the 
slopes of the measured species and we conclude that this result 
suggests a significant contribution of non-thermal processes (such as mini-jets) 
to the particle spectra.

\section{Acknowledgments}

We would like to thank S. Albino, W. Vogelsang, and K. Werner 
for several illuminating discussions and for the calculations they have 
provided.
We thank the RHIC Operations Group and RCF at BNL, and the
NERSC Center at LBNL for their support. This work was supported
in part by the HENP Divisions of the Office of Science of the U.S.
DOE; the U.S. NSF; the BMBF of Germany; IN2P3, RA, RPL, and
EMN of France; EPSRC of the United Kingdom; FAPESP of Brazil;
the Russian Ministry of Science and Technology; the Ministry of
Education and the NNSFC of China; IRP and GA of the Czech Republic,
FOM of the Netherlands, DAE, DST, and CSIR of the Government
of India; Swiss NSF; the Polish State Committee for Scientific 
Research; STAA of Slovakia, and the Korea Sci. \& Eng. Foundation.


\newpage

\begin{thebibliography}{99}

\bibitem{Bocquet96}G. Bocquet {\it et al.}, Phys. Lett. B {\bf 366}, 441 (1996).
\bibitem{Alper75}B. Alper {\it et al.} (British-Scandinavian Collaboration), Nucl. Phys. B {\bf 100}, 237 (1975).
\bibitem{E735}T. Alexopoulos {\it et al.} (E735 Collaboration), \prd {\bf 48}, 984 (1993).
\bibitem{E735-2}T. Alexopoulos {\it et al.} (E735 Collaboration), Phys. Lett. B {\bf 336}, 599 (1994).
\bibitem{Gyulassy92}X. N. Wang and M. Gyulassy, Phys. Lett. B {\bf 282}, 466 (1992).
\bibitem{STAR}K. H. Ackermann {\it et al.} (STAR Collaboration), Nucl. Instrum. Meth. A {\bf 499}, 624 (2003).
\bibitem{TPC}M. Anderson {\it et al.}, Nucl. Instrum. Meth. A {\bf 499}, 659 (2003).
\bibitem{Bichsel}H. Bichsel, Nucl. Instrum. Meth. A {\bf 562}, 154 (2006); H. Bichsel, D. E. Groom, S. R. Klein, J. Phys. G {\bf 33}, 258 (2006).
\bibitem{ppSpecPaper}J. Adams {\it et al.} (STAR Collaboration), \prl {\bf 92}, 112301 (2004).
\bibitem{STARBBCs}J. Adams {\it et al.} (STAR Collaboration), \prc {\bf 70}, 041901 (2004). 
\bibitem{CTB}F. S. Bieser {\it et al.}, Nucl. Instrum. Meth. A {\bf 499}, 766 (2003).
\bibitem{STAR130Mult}C. Adler {\it et al.} (STAR Collaboration), \prl {\bf 87}, 262302 (2001).
\bibitem{STAR130p-bar-p}C. Adler {\it et al.} (STAR Collaboration), \prl {\bf 86}, 4778 (2001); {\it ibid} {\bf 90}, 119903 (2003).
\bibitem{UA5}R. E. Ansorge {\it et al} (UA5 Collaboration), Nucl. Phys. B {\bf 328}, 36 (1989).
\bibitem{UA5:87}R. E. Ansorge {\it et al} (UA5 Collaboration), Phys. Lett. B {\bf 199}, 311 (1987).
\bibitem{Hagedorn}R. Hagedorn, Riv. Nuovo Cimento {\bf 6}, n. 10, 1 (1984).
\bibitem{Wilk}G. Wilk and Z. W\l{}odarczyk, \prl {\bf 84}, 2770 (2000).
\bibitem{UA5:546}G. J. Alner {\it et al.} (UA5 Collaboration), Phys. Rep. {\bf 154}, 247 (1987).
\bibitem{pythia}T. Sj\"{o}strand {\it et al.}, Computer Physics Commun. {\bf 135}, 238 (2001).
\bibitem{ISR75}B. Alper {\it et al.}, Nucl. Phys. B {\bf 87}, 19 (1975).
\bibitem{ISR81}K. Alpgard {\it et al.}, Phys. Lett. B {\bf 107}, 310 (1981).
\bibitem{Wong}G. Gatoff and C. Y. Wong, \prd {\bf 46}, 997 (1992).
\bibitem{Schaffner-Bielich02}J. Schaffner-Bielich, D. Kharzeev, L. McLerran, and R. Venugopalan, nucl-th/0202054
\bibitem{Schaffner-BielichCGC02}J. Schaffner-Bielich, D. Kharzeev, L. McLerran, and R. Venugopalan, Nucl. Phys. A {\bf 705}, 494 (2002).
\bibitem{STAR-pi-k-p}J. Adams {\it et al.} (STAR Collaboration), \prl {\bf 92}, 112301 (2004).
\bibitem{STAR-TOF}J. Adams {\it et al.} (STAR Collaboration), Phys. Lett. B {\bf 616}, 8 (2005).
\bibitem{STAR-rDEDX}J. Adams {\it et al.} (STAR Collaboration), Phys. Lett. B {\bf 637}, 161 (2006).
\bibitem{phenix-pi0}S. S. Adler {\it et al.} (PHENIX collaboration), \prl {\bf 91}, 241803 (2003).
\bibitem{HarrisSQM03}J. Harris, J. Phys. G {\bf 30}, S613 (2004).
\bibitem{HERA}A. D. Martin, W. J. Stirling, and R. G. Roberts, \prd {\bf 50}, 6734 (1994).
\bibitem{Xin-Nian}X. N. Wang, \prc {\bf 58}, 2321 (1998).
\bibitem{ISRcurve}M. Bourquin and J. M. Gaillard, Nucl. Phys. B {\bf 114}, 334 (1976).
\bibitem{ISR25GeV}B. Alper {\it et al.}, Nucl. Phys. B {\bf 114}, 1 (1976).
\bibitem{FNAL}J. W. Cronin {\it et al.} (Chicago-Princeton Group), \prd {\bf 11}, 3105 (1975).
\bibitem{Dumitru}A. Dumitru and C. Spieles, Phys. Lett. B {\bf 446}, 326 (1999).
\bibitem{Gyu92_2}X. N. Wang and M. Gyulassy, \prd {\bf 45}, 844 (1992).
\bibitem{string1}T. Sj\"{o}strand, hep-ph/9508391; B. Andersson, G. Gustafson, G. Ingelman, 
and T. Sj\"{o}strand, Phys. Rep. {\bf 97}, 31 (1983).
\bibitem{string2}B. Andersson, ``The Lund Model" (Cambridge University Press, 1998); B. Andersson {\it et al.}, hep-ph/0212122.
\bibitem{Field} R. D. Field, \prd {\bf 65}, 094006 (2002); R. D. Field, hep-ph/0201192.
\bibitem{kfactor} R. Vogt, Heavy Ion Phys. \textbf{17}, 75 (2003).
\bibitem{eskola-kfactor} K. Eskola and H. Honkanen, Nucl. Phys. A \textbf{713}, 167 (2003).
\bibitem{Lamont06}J. Adams {\it et al.} (STAR Collaboration), nucl-ex/0601042, submitted to Phys. Rev. C.
\bibitem{KKP} B. A. Kniehl, G. Kramer, and B. P\"otter, Nucl. Phys. B {\bf 597}, 337 (2001).
\bibitem{Vogelsang} D. deFlorian, M. Stratmann, and W. Vogelsang, \prd \textbf{57}, 5811 (1998).
\bibitem{STAR-pi0} J. Adams \textit{et al.} (STAR Collaboration), \prl {\bf 92}, 171801 (2004).
\bibitem{Bourrely Soffer} C. Bourrely and J. Soffer, \prd \textbf{68}, 014003 (2003).
\bibitem{OPAL}G. Abbiendi \textit{et al.} (OPAL Collaboration), Eur. Phys. J. C 16, 407 (2000).
\bibitem{AKK} S. Albino, B. A. Kniehl, and G. Kramer, Nucl. Phys. B {\bf 725}, 181 (2005). 
\bibitem{Fai}X. Zhang, G. Fai and P. Levai, \prl \textbf{89}, 272301 (2002).
\bibitem{FPD}J. Adams {\it et al.} (STAR Collaboration), nucl-ex/0602011, submitted to Phys. Rev. Lett.
\bibitem{EPOSDetails} K. Werner, F. M. Liu, and T. Pierog, hep-ph/0506232.
\bibitem{EPOS-RHIC} K. Werner, F. M. Liu, and T. Pierog, J. Phys. G {\bf 31}, S985 (2005). 
\bibitem{Hagedorn65}R. Hagedorn, Nuovo Cimento Suppl. {\bf 3}, 147 (1965).
\bibitem{Hagedorn68a}R. Hagedorn and J. Randt, Nuovo Cimento Suppl. {\bf 6}, 169 (1968).
\bibitem{Hagedorn68b}R. Hagedorn, Nuovo Cimento Suppl. {\bf 6}, 311 (1968).
\bibitem{Hagedorn70}R. Hagedorn, Nucl. Phys. B {\bf 24}, 93 (1970).
\bibitem{Siemens79}P. J. Siemens and J. I. Kapusta, \prl {\bf 43}, 1486 (1979).
\bibitem{Mekjian82}A. Z. Mekjian, Nucl. Phys. A {\bf 384}, 492 (1982).
\bibitem{Csernai86}L. Csernai and J. Kapusta, Phys. Rep. {\bf 131}, 223 (1986).
\bibitem{Stocker86}H. St\"ocker and W. Greiner, Phys. Rep. {\bf 137}, 277 (1986).
\bibitem{Cleymans93}J. Cleymans and H. Satz, Z. Phys. C {\bf 57}, 135 (1993).
\bibitem{Rafelski02}J. Rafelski and J. Letessier, J. Phys. G {\bf 28}, 1819 (2002).
\bibitem{PBM03}P. Braun-Munzinger, K. Redlich, and J. Stachel, nucl-th/0304013.
\bibitem{Becattini}F. Becattini and U. Heinz, Z. Phys. C {\bf 76}, 269 (1997).
\bibitem{Becattinie+e-}F. Becattini, A. Giovannini, and S. Lupia, Z. Phys. C {\bf 72}, 491 (1996).
\bibitem{Biro}T. Biro and B. M\"uller, Phys. Lett. B {\bf 578}, 78 (2004).
\bibitem{Hormuzdiar}J. Hormuzdiar, S. D. H. Hsu, and G. Mahlon, Int. J. Mod. Phys. E {\bf 12}, 649 (2003).
\bibitem{THERMUS}S. Wheaton and J. Cleymans, hep-ph/0407174.
\bibitem{Liu04}F. M. Liu, J. Aichelin, M. Bleicher, and K. Werner, \prc {\bf 69}, 054002 (2004).
\bibitem{Liu03}F. M. Liu, K. Werner, and J. Aichelin, \prc {\bf 68}, 024905 (2003).

\end{thebibliography}
\end{document}